\documentclass[useAMS,usenatbib]{mn2e}

\usepackage{amsmath,graphicx,wrapfig,lipsum,caption,overpic,amssymb}
\usepackage{subfigure}

\newcommand{\ignore}[1]{}

\def\mdot{\dot{M}}

\title[Composition of Early Planetary Atmospheres II]{Composition of Early Planetary Atmospheres II: Coupled Dust and Chemical Evolution in Protoplanetary Disks}
\author[A.J. Cridland et al.]{A.J. Cridland$^{1}$\thanks{E-mail:
cridlaaj@mcmaster.ca}, Ralph E. Pudritz$^{1,2,3,4}$\thanks{E-mail:
pudritz@mcmaster.ca}, Tilman Birnstiel$^5$, 
\newauthor L. Ilsedore Cleeves$^6$ \& Edwin A. Bergin$^7$ \\
$^{1}$Department of Physics and Astronomy, McMaster University, Hamilton, Ontario, Canada, L8S 4E8 \\ $^2$Origins Institute, McMaster University, Hamilton, Ontario, Canada, L8S 438 \\ $^3$Zentrum f\"ur Astronomie der Universit\"at Heidelberg, Institut f\"ur Theoretische Astrophysik, Albert-Ueberle-Str. 2, 69120 Heidelberg, Germany \\ $^4$Max Planck Institute for Astronomy K\"onigstuhl 17, D-69117 Heidelberg, Germany \\ $^5$University Observatory, Faculty of Physics, Ludwig-Maximilians-Universit\"at M\"unchen, Scheinerstr. 1, 81679 Munich, Germany
\\ $^6$Harvard-Smithsonian Cneter for Astrophysics, 60 Garden Street, Cambridge, MA 02138, USA \\ $^7$Department of Astronomy, University of Michigan, 1085 S. University Avenue, Ann Arbor, MI 48109, USA}

\begin{document}
\bibliographystyle{mn2e}
\date{\today}

\pagerange{\pageref{firstpage}--\pageref{lastpage}} \pubyear{2015}

\maketitle

\label{firstpage}

\begin{abstract}

We present the next step in a series of papers devoted to connecting the composition of the atmospheres of forming planets with the chemistry of their natal evolving protoplanetary disks.  The model presented here computes the coupled chemical and dust evolution of the disk and the formation of three planets per disk model. Our three canonical planet traps produce a Jupiter near 1 AU, a Hot Jupiter and a Super-Earth. We study the dependency of the final orbital radius, mass, and atmospheric chemistry of planets forming in disk models with initial disk masses that vary by 0.02 $M_\odot$ above and below our fiducial model ($M_{disk,0} = 0.1 ~M_\odot$). We compute C/O and C/N for the atmospheres formed in our 3 models and find that C/O$_{\rm planet}\sim$ C/O$_{\rm disk}$, which does not vary strongly between different planets formed in our model. The nitrogen content of atmospheres can vary in planets that grow in different disk models. These differences are related to the formation history of the planet, the time and location that the planet accretes its atmosphere, and are encoded in the bulk abundance of NH$_3$. These results suggest that future observations of atmospheric NH$_3$ and an estimation of the planetary C/O and C/N can inform the formation history of particular planetary systems.
\end{abstract}

\begin{keywords}
protoplanetary discs, interplanetary medium
\end{keywords}

\section{ Introduction }\label{sec:intro}

What physical processes set the observable chemical abundances in exoplanetary atmospheres? The bulk abundance of the atmosphere will ultimately depend on the chemical structure of the gas that is present in the planet's natal protoplanetary disk \citep{BerOber16,Madhu14,Madhu16,Mordasini16}. The disk astrochemistry depends on the physical state of the disk, which includes the temperature, density, and ionization of the gas, and radiative flux of high energy photons \citep{Fogel11,Helling14}. In this and past works \citep{Alessi16,Crid16a,Crid16b} we present a model of an evolving astrochemical disk which when combined with the core accretion model for planetary formation \citep{IL04a} can predict the bulk chemical abundance of planetary atmospheres.

An `early' planetary atmosphere is a young ($t_{age} \lesssim 10$ Myr) atmosphere that has not had enough time to physically or chemically evolve to change its observable chemical signature. As such, we do not compute the physical or chemical structure of the atmosphere as it accretes its gas. Instead our work focuses on computing the bulk chemical abundance of the atmosphere that arise from the chemical abundance of the disk's gas that the planet accretes.

The available observational data from secondary eclipse spectroscopy (ie. \cite{Line14}, \cite{Lee12}, \cite{Moses13}) and transit spectroscopy (ie. \cite{Ranjan14}, \cite{Sing15}, \cite{Wakeford15}) have begun to characterize the chemical composition of exoplanetary atmospheres, detecting species like H$_2$O, CO$_2$, CO, CH$_4$, K, and Na for about a dozen planets. Early connections between theoretical work and observations identified the carbon-to-oxygen elemental ratio (C/O) as an excellent candidate for linking the planet formation process and the resulting atmospheric chemistry \citep{Oberg11,Madu2014,Thiabaud2015}. This ratio varies with disk radius as volatiles are either frozen out or sublimated into the gas phase. Hence C/O of a planet's atmosphere will depend on where that planet accreted its gas, relative to the ice line of volatiles like H$_2$O, CO$_2$, and CO \citep{Oberg11}. 

A recent synthetic model that calculated the links between  the formation and migration of hot Jupters in evolving disks,  to the construction and composition of their  atmospheres used simplified models of the composition of planetesimals \citep{Mordasini16}.  The resulting C/O ratios in this analysis were subsolar because of the O rich nature of the accreting solid materials.  

In this paper we introduce a new method of computing the coupled chemical and dust evolution of the disk, in order to give a first combined treatment of these two critical aspects of disk evolution and early planetary formation. We use this approach to then address key problems in linking planet formation with the resulting composition of planetary atmospheres. One application of this approach is to demonstrate that the famous C/O and the carbon-to-nitrogen ratio (C/N) may only weakly dependent on the formation history of planets. And that the particular abundance of nitrogen carrying molecules trace details in the planet's formation history like the timing of the planet's gas accretion \citep{Crid16a}.

Of particular interest to our work is whether the bulk chemical compositions of these `early' planetary atmospheres can model the bulk compositions that are inferred by emission and transmission spectra of observed planets. In doing so we might learn from where the diversity of atmospheric chemistry is inherited.

The paper is structured as follows: in Section \S \ref{sec:back} we present background information from our previous works, and in Section \S \ref{sec:model} we outline our physical model including our determination of the location of the dead zone. In Section \S \ref{sec:CDG} we show how the different dust models change the distribution of gas species in the disk, while in \S \ref{sec:res01} we report the results of the planet formation in the fiducial model of \cite{Crid16b} (CPB17) and compare the results to those of \cite{Crid16a} (CPA16). In Section \S \ref{sec:res02} we report the results of the two supplementary disks and compare their resulting planets to the fiducial model. In Section \S \ref{sec:dis} we discuss the implications of our results, particularly how a population synthesis model will improve our understanding of variations in atmospheric chemical abundances. We finish with concluding remarks in Section \S \ref{sec:con}.

\section{ Background }\label{sec:back}

Our model involves the combination of accretion disk physics, dust physics (coagulation, fragmentation, radial drift and settling), radiative transfer, astrochemsitry (photochemistry, gas phase, and grain surface) and planet formation theory (core accretion and migration). A particularly important aspect of the planet formation model is the use of `planet traps' to set the location of the protoplanet and limit its rate of migration. Planet traps are inhomogeneities in disk properties where the net torque vanishes, so that migration occurs at the much slower rate of disk evolution. Introduced by \cite{Masset06} to investigate planetary dynamics at the inner edge of a disk dead zone, planet traps have been generalized and used in population synthesis models to replicate the number of planets observed in different regions of the mass-semi-major axis diagram (for ex. \cite{HP11,HP12,HP13}). 

A planet trap works because of the relative strength of the Lindblad and co-rotation torques \citep{Masset06}. The Lindblad torque is caused by the gravitational pull of material that is perturbed from Lindblad resonances near the planet back upon the planet \citep{Ward1997}. The two nearest resonances are at radial positions with orbital frequencies that are twice the planet's (outer resonance) and half the planet's (inner resonance) orbital frequency. Generally the outer resonance is stronger than the inner resonance and the planet migrates to smaller orbital radii.

The co-rotation torque is caused by gas with orbital frequencies that are close the planet's. The gas enters into horseshoe orbits where they oscillate between orbital radii that are smaller than the planet's to orbital radii that are larger. At the turn of each horseshoe orbit the gas exchanges angular momentum with the planet \citep{Paard10}. If each turn is symmetrical the co-rotation torque does not exert a net torque on the planet - this is called saturation. With enough turbulence the gas mixes with gas on the opposite side of the horseshoe orbit which causes an asymmetry in the two turns and a net torque which generally causes outward migration \citep{Seager10,Paard10}. Outside of planet traps, the co-rotation torque is generally weaker than the Lindblad torque.

In our work we are interested in three planet traps: the water `ice line', the heat transition and the dead zone. The water ice line depends on the chemical state of the gas and is located where the water content of the disk transitions from being primarily in the vapour phase to being in the ice phase. The heat transition traces the location where the disk's heating mechanism transitions from being primarily due to viscous stresses to direct irradiation from the host star. Finally the dead zone is located at the transition point between a low ionization state where the disk turbulence is weak (`dead'), to a high ionization state where the turbulence is active \citep{MP09}.

Our fiducial model (CPA16) includes all of the above effects, but assumed a constant dust to gas ratio at all times throughout the disk. In that work we show that the chemical abundance of planetary atmospheres depends on both {\it where} and {\it when} the planet accretes its gas. We observe that C/O and C/N ratio encode information regarding the formation history of the planet. Atmosphere C/O ratios that were nearly the same as the disk's initial C/O ratio were found for planets that formed within the water condensation front (ice line) of the disk. We expect that higher C/O ratios (C/O $\sim$ 1, compared to solar: 0.54 and our initial disk: 0.288) will be found in planets that accrete their atmosphere outside the ice line, however this was not demonstrated in that previous paper. The sub-solar C/O of our initial disk was due to our choice of the initial molecular species that are inherited by the disk. Our initial abundances are 
based on the work of \cite{Fogel11} and \cite{AH99} and represents the chemical state of a dense molecular cloud. 

The C/N ratio is encoded {\it when} the planet accretes its atmosphere. Atmospheres that were accreted early ($t \lesssim 1$ Myr) have higher C/N ratio by almost a factor of 5 over planets that accrete their atmosphere later ($t > 1$ Myr). The temperature of the disk's gas during accretion is what leads to the large change in C/N between the early and late accretors. Early on, the inner disk ($r \lesssim 5$ AU) is warmer due to viscous heating and the nitrogen content is primarily as NH$_3$ and HCN which leads to higher C/N than later when the disk is cooler and nitrogen is found primarily as N$_2$.

We incorporated the effects of dust coagulation, fragmentation, and radial drift to our model in CPB17 and demonstrated that the gas disk became more ionized when compared to the simpler dust model of CPA16 because the radial drift and growth of the grains reduced the disk's opacity to high energy radiation. The ionization structure of the disk is important to our work because it sets the location of one of the planet traps: the dead zone. We assume that the source of turbulence in the disk is driven by the magnetorotational instability (MRI) which is caused by a coupling of the disk's magnetic field to free electrons in the gas (for ex. \cite{BalbusHawley91}). The growth of turbulence due to the MRI is counteracted by the diffusion of the magnetic field. We assume that the primary source of diffusion on the midplane of the disk is Ohmic resistivity \citep{BaiStone2013,Gressel2015}. This assumption ignores other non-ideal effects like ambipolar diffusion \citep{BaiStone2013,Gressel2015} and the Hall effect \citep{BaiStone2016}. The impact of these other sources of diffusion are more complicated to estimate and are part of an ongoing discussion regarding the turbulent structure of protoplanetary disks.

In this paper, we compute the formation of the three planets and the chemical structure of their atmospheres in the resulting disk model from CPB17. We will compare the results of our work to the results of CPA16 where we use the same disk parameters, but change the way that the dust physics is handled. Importantly, the more complicated dust physics shows that the disk's dead zone evolved radially much faster than in the disk model with a constant dust to gas ratio without grain coagulation or fragmentation. The faster evolution causes the planet forming in the dead zone trap to have a different formation history, and potentially a different atmospheric composition. In CPA16 the dead zone planet failed to accrete an atmosphere in the 4.1 Myr disk lifetime.

With the need for turbulence to maintain an unsaturated co-rotation torque, the relative location of the dead zone edge and other planet traps dictate whether the planet remains trapped over its formation lifetime. Within the dead zone, we assume that the turbulent $\alpha$ drops by two orders of magnitude when compare to the active region. This reduction leads to a lengthened eddy turnover time, and a saturated corotation torque if the eddy turnover time exceeds the libration time of the horseshoe orbit (see for example CPA16). In CPA16 we found that the planet forming in the heat transition trap saturated 1.1 Myr into formation, causing the planet to quickly migrate and become a Hot Jupiter. In CPB17 we showed that the dead zone was located at smaller radii than the heat transition trap by 1.3 Myr. If the dead zone crosses the heat transition trap's location the co-rotation will remain unsaturated and the planet will remain trapped for its entire formation history. This will produce a different planet than 
the one reported in CPA16.

Finally it was pointed out in CPA16 that to fully understand the connection between the atmospheric abundance predictions and our model, we would require a statistically significant sampling of possible disk parameters like initial mass and lifetime. In \cite{Alessi16} it was shown that both of these disk parameters impact the formation history of planets in our formation model. In particular the initial mass of the disk changes the location and evolution of the planet traps, hence the location of the forming planet. Ideally, a full population synthesis model (ie. \cite{IL04a,HP13}; Alessi \& Pudritz (in prep.)), with values drawn from distributions of a few parameters, would be applied in our evolving astrochemical disk. In preparation for a future complete population synthesis study, we report results from two additional disk models where we have varied the initial disk mass by 0.02 $M_\odot$ above and below our fiducial disk of 0.1 $M_\odot$.

\section{ Physical Model }\label{sec:model}

Our evolving astrochemical model as well as our method of identifying the planet trap locations is outlined in detail in Cridland, Pudritz \& Alessi (2016). Our dust model is outlined in Cridland, Pudritz, Birnstiel (2017)  and was based on the model of \cite{B12}. Here we outline the important aspect of each of these models.

Generally we compute the temperature and surface density radial distribution of the gas with an analytic solution of the gas diffusion equation (see \cite{Cham09}, CPA16). This self-similar model for disk structure and evolution assumes that the heating of the gas is dominated by either viscous evolution or direct irradiation of the host star. Furthermore, as a computation constraint we assume that the disk's opacity and $\alpha$-parameter (ie. in \cite{SS73}, due to the presence of disk winds and/or turbulence) are constant in space and time, and that the mass accretion rate is constant in space but evolves with time (see \cite{Alessi16} and CPA16 for details).

This gas model is combined with a model for the evolution of the dust surface density \citep{B12}. The particle size distribution is then reconstructed with the semi analytical treatment of \cite{B15}. The dust and gas distributions are computed over many `snapshots' over a disk lifetime of 4.1 Myr. 

In each snapshot we compute the flux of UV and X-ray photons with RADMC3D, a Monte Carlo radiative transfer scheme \citep{RADMC}. The wavelength dependent disk opacity is computed by the on-the-fly method in RADMC3D and depends on the dust surface density and size distribution using optical constants from \cite{Draine03}. We compute the flux of 4 UV and 3 X-ray test wavelengths, then extrapolate their results for the wavelengths: 930 - 2000 angstroms for UV and 1 - 20 keV for X-ray using sample T Tauri UV \citep{Fogel11,Bethell11} and X-ray \citep{Kastner99,Kastner02,Cleeves15} spectra as guides.

We note that relying solely on dust as our source of disk opacity does ignore the X-ray cross section of the gas. Over our range of wavelengths dust does dominate the X-ray cross section \citep{Bethell11}, however we will likely underestimate the disk's opacity late in it's evolution as the dust surface density drops. At these late times, the gas will begin to dominate the cross section and set a floor to the disk's opacity.

Next we compute the astrochemistry using a chemical kinetic code (ie. \cite{Fogel11,Cleeves14}, CPA16) in each snapshot using the gas, dust, and radiation as inputs. This process produces an evolving astrochemical disk in which we computed the formation of a planet using the core accretion model and the migration of the forming planet with a planet trap model. Some specific features of this theoretical framework are discussed below.

\subsection{ Evolving Astrochemical Model }\label{ssec:EAM}

An important feature in our astrochemical model is the fact that the temperature and density profiles evolve with time as the disk ages. The chemistry is handled with a non-equilibrium code from \cite{Fogel11} and \cite{Cleeves14}. This astrochemical code includes, among others, ion-driven gas phase reactions, the freeze out of volatiles, photo-desorption, and the production of water through grain surface reactions. In the chemical model we compute an average grain size based on the results of our dust model (see below), and allow the average grain size to vary at different disk radii and evolve in time.

We combined this chemical model with an analytic disk model from \cite{Cham09} which models a disk that evolves due to mass accretion caused by viscous stresses. This analytic model produces temperature and surface density profiles that scale as:\begin{align}
T \propto R^\beta ;\quad \Sigma\propto R^s,
\label{eq:mod01}
\end{align}
where the exponents $\beta$ and $s$ are determined by the source of heating. Viscous heating and direct irradiation are the primary heating sources. Viscous heating is caused by gravitational energy being released as material accretes through the disk. This heating scales with the mass accretion rate $\mdot$ which is reduced as time passes. Because of this dependence, the region of the disk that is viscously heated will cool as the disk ages. At larger radii the densities are lower and the heating is dominated by the direct irradiation of the disk from the host star. We assume that the stellar properties of the host star do not change, so the temperature profile of the disk is also static. This assumption ignores the effect of excess high energy photons due to the accretion shock on the heating of the disk. In doing so we have likely underestimated the temperature of the disk in the irradiated region at the earliest times. However, as the disk ages and the accretion rate drops the heating from the accretion 
shock is drastically reduced. In a future work we will explore the impact of both the temperature structure and chemistry of an evolving stellar emission spectrum.

Viscous heating dominates at smaller radii where the densities are the highest, but as the disk accretes its density and mass accretion rate both drop. As the mass accretion rate drops, the region of the disk where viscous heating dominates shrinks and the point where the temperature from viscous heating is the same as the temperature from irradiation (the heat transition trap) moves to smaller radii.

The evolving viscously heated region of the disk also has important implications to the chemistry in the disk. Features like the water ice line, which closely depends on the temperature of the gas, will evolve to lower radii as the disk cools. Additionally the main molecular carrier of some elements, like nitrogen, can change depending on the temperature of the gas. In hotter disks the nitrogen will generally be found in NH$_3$ and HCN, while in cooler disks it is mostly found in N$_2$ (CPA16). This change in elemental carrier will be imprinted in the bulk chemical abundance of the planetary atmosphere based on {\it where} and {\it when} the planet accretes its gas.

Just as the temperature profile evolves, the surface density profile of both the gas and dust also changes with time. The gas density drops as mass is accreted onto the host star due to turbulent viscous stresses and magnetocentrifugal winds. Likewise the dust also accretes onto the host star through viscous stresses, however the it is also susceptible to a faster source of evolution: radial drift. Radial drift involves the sub-Keplarian orbit of dust grains caused by a gas drag which scales with the surface area of the grain. As a result the largest grains are most affected by radial drift, which when combined with coagulation causes a faster reduction in surface density than viscous evolution alone. The dependency of the radial drift rate with grain surface area extends up to a point where the grain's inertia becomes too great for gas drag to have a large impact on the grain's orbit. Our size distribution extends up to a few tens of cm, and so we do not simulate the dust evolution up to this cut off. The 
growth and drift of dust grains lead to an overall reduction in the disk's opacity to high energy photons, and an increase in the ionization of the gas (CPB17). While we do evolve the radial drift of dust grains (see below), we ignore the transport of frozen volatiles on the grains as they migrate inwards. In principle, grains will bring volatiles from the outer disk to smaller radii as they drift which could enhance the abundance of volatiles in the planet-forming regions of the disk (r $\lesssim 10$ AU). We leave an investigation into this effect to future work.

Ions drive chemical reactions because of their lower activation barrier when compared to neutral-neutral gas reactions. As mentioned above they also cause turbulence through the MRI when coupled to the disk's magnetic field. Because the chemical and physical state of the gas is sensitive to the flux of ionizing photons, an accurate description of disk opacity and evolution of the dust is an important feature of our model.

We use the same initial chemical abundances as in CPA16, derived from \cite{Fogel11} and \citep{AH99}, see Table \ref{tab:initchem}.

\begin{table}
\centering
\caption{Initial abundances relative to the number of H atoms. Included is the initial ratio of carbon atoms to oxygen atoms (C/O) and the initial ratio of carbon atmos to nitrogen (C/N).}\label{tab:initchem}
\begin{tabular}{l l l l}\hline\hline
Species & Abundance & Species & Abundance \\\hline
H$_2$ & $0.5$ & H$_2$O & $2.5\times 10^{-4}$ \\
He & $0.14$ & N & $2.25\times 10^{-5}$ \\
CN & $6.0\times 10^{-8}$ & H$_3^+$ & $1.0\times 10^{-8}$ \\
CS & $4.0\times 10^{-9}$ & SO & $5.0\times 10^{-9}$ \\
Si$^+$ & $1.0\times 10^{-11}$ & S$^+$ & $1.0\times 10^{-11}$ \\
Mg$^+$ & $1.0\times 10^{-11}$ & Fe$^+$ & $1.0\times 10^{-11}$ \\
C$^+$ & $1.0\times 10^{-9}$ & GRAIN & $6.0\times 10^{-12}$ \\
CO & $1.0\times 10^{-4}$ & N$_2$ & $1.0\times 10^{-6}$ \\
C & $7.0\times 10^{-7}$ & NH$_3$ & $8.0\times 10^{-8}$ \\
HCN & $2.0\times 10^{-8}$ & HCO$^+$ & $9.0\times 10^{-9}$ \\
C$_2$H & $ 8.0\times 10^{-9}$ & C/O & $0.288$ \\
& & C/N & $4.09$ \\
\hline
\end{tabular}
\end{table} 

\subsection{ Dust Model }\label{ssec:DM}

Our evolving dust model was presented in CPB17 and is based on the Two-population model presented in \cite{B12}. This model involves estimating the size and density distribution of the dust grains by computing the evolution of two sample dust populations, one representing the smallest grains (monomers) and one representing the large population. The size distribution is assumed to be a power law with a monomer size of 0.1 microns and a maximum size set by either fragmentation or radial drift depending on which physical process has the lower timescale.

This simplified dust model has been tested against full coagulation simulations, reproducing the general trends from those more complicated models \citep{B12}. A benefit of the Two-population model is how quickly the evolution can be computed. Depending on the treatment of the fragmentation, standard dust evolution codes scale $O(N^2)$ where $N$ is the number of grain sizes used in the simulation. So the Two-population model, which computes the coagulation, fragmentation and radial drift of only two populations can be run much faster than a standard coagulation simulations that is resolved with at least 100 different sizes.

Our dust model includes the effect of a changing fragmentation threshold speed across the water ice line, as first discussed in \cite{B10}. The fragmentation threshold speed is the minimum relative collision speed that results in the fragmentation of the grains. This threshold speed depends on the chemical nature of the dust grain, its porosity, and the amount of ice coverage. For simplicity we assume a threshold speed of 10 m/s outside the water ice line because the grain is fully covered, and hence strengthened by a layer of ice \citep{Wada09,GundlachBlum2015}. Within the ice line the layer of ice is gone and the threshold speed is reduced to 1 m/s. We used the chemical results of our fiducial model (CPA16) to fit the location and evolution of the water ice line as a function of time. Using this fit we allowed the location of the water ice line to evolve throughout the evolution of the dust.

The impact of this changing fragmentation threshold speed leads to a reduction in the size of the largest grain by about two orders of magnitude within the water ice line (CPB17, or see Figure \ref{fig:res04} below). These smaller grains radially drift slower than larger ones, with a drift time scale of the form \citep{B10}:\begin{align}
\tau_{drift} &= \frac{r V_k}{St c_s^2}\gamma^{-1}
\label{eq:drift}
\end{align}
at the radial position $r$. The Kepler speed ($V_k$) and gas sound speed ($c_s$) are evaluated at the position $r$. The Stokes number scales linearly with the size of the grain and encodes the coupling between the grain and the gas. Longer drift time scales within the ice line lead to an enhancement of the dust surface density of an order of magnitude within the water ice line. This enhancement has two important affects on planet formation: first it shields the inner disk from high energy photons, reducing the level of ionization early on in the disk lifetime. Secondly, the enhancement will increase the growth rate of the forming planet because there is more material available during oligarchic growth.

\subsection{ Planet Formation Model }\label{ssec:PF}

Our planet formation model relies on the Core Accretion model as seen in \cite{IL04a} to compute the rate of mass accretion onto the planet. This core accretion model is used in conjunction with the planet trap model that was outlined above. The combination of these two models has been used in population synthesis models to reproduce the population statistics that have been observed in the exoplanetary data (\cite{HP13}; Alessi \& Pudritz (in prep.)).

\subsubsection{ Core Accretion Model }\label{sssec:CA}

The Core Accretion model contains three important phases of accretion. The first is known as oligarchic growth, where a single core accretes solid material from a sea of 10-100 km sized planetesimals. We do not model the growth of these planetesimals from micron sized dust grains up to their assumed sizes, instead we assume that all of the available solid material is in planetesimals. In doing this, we neglect the reduction in the disk's opacity caused by the rapid agglomeration of dust particles caused by the streaming instability (see for ex. \cite{Schafer17}). The growing planet will impact the surface density of the surrounding gas and dust, and as a result will locally change the temperature and ionization of the gas. These effects are beyond the scope of this work.

The rate of accretion is fast, with timescales $O(10^5)$ yr. The solid accretion rate has the form (\cite{IL04a}, their equations 5 and 6): \begin{align}	
\frac{dM_c}{dt} = & \frac{M_c}{\tau_{c,acc}} \nonumber\\
	\simeq & \frac{M_c}{1.2\times 10^5} \left(\frac{\Sigma_d}{10 {\rm gcm}^{-2}}\right)\left(\frac{a}{1 {\rm AU}}\right)^{-1/2} \nonumber\\
	\times & \left(\frac{M_c}{M_\oplus}\right)^{-1/3}\left(\frac{M_s}{M_\oplus}\right)^{1/6} \nonumber\\
		\times & \left[\left(\frac{\Sigma_g}{2.4\times 10^3 {\rm gcm}^{-2}}\right)^{-1/5} \left(\frac{a}{1 {\rm AU}}\right)^{1/20}\left(\frac{m}{10^{18} {\rm g}}\right)^{1/15}\right]^{-2} \nonumber\\
	& \quad {\rm g~yr}^{-1}.
\label{eq:acc01}
\end{align}

The growing core quickly accretes all of the solid material in its immediate vicinity. At this point it has reached its `isolation mass' \citep{IL04a}, which represents the maximum available material to the planet if the planet does not migrate. Once it has reached its isolation mass the rate of planetesimal accretion drops, allowing for gas accretion to take place.

We assume that the gas accretion rate is limited only by the Kelvin-Helmholtz timescale which we model with \begin{align}
\tau_{KH} \simeq 10^c \left(\frac{M_{plnt}}{M_\oplus}\right)^{-d}~ {\rm yr},
\label{eq:mod02}
\end{align}
so that the mass of the planet grows as\begin{align}
\frac{dM_{plnt}}{dt} = \frac{M_{plnt}}{\tau_{KH}},
\label{eq:mod03}
\end{align} 
for further details see \cite{Alessi16}. The form of equation \ref{eq:mod02} has been demonstrated by numerical simulations, and the constants $c$ and $d$ have been shown to have values between 8-10 and 2-4 respectively \citep{IL04a}. We follow \cite{IL04a} in choosing $c=9$ and $d=3$. This choice was also used in \cite{HP13} to recreate the distribution of planets on the mass semi-major axis diagram derived from observations. 

While in practice the gas surface density is locally reduced by the accreting planet (see for example \cite{Ormel2015}), we do not update the disk gas density and temperature while the planet draws down its atmosphere.

\subsubsection{ Planetary Migration }\label{sssec:PT}

As it forms, the protoplanet and its natal accretion disk exchange angular momentum through the Lindblad and co-rotation torques \citep{Paard14} which generally results in a loss of angular momentum for the protoplanet. This loss causes the planet to `migrate' to smaller orbital radii in a process called Type-I migration.

As was previously stated, Type-I migration is limited by discontinuities in disk properties known as `planet traps' \citep{Masset06}. Here we assume that a trapped planet will exactly follow the radial location of the its planet trap. If the Ice line and Heat transition traps are found within the Dead zone, we check at every timestep whether the co-rotation torque saturates by comparing the eddy turnover time of the gas to the gas libration time. If the libration time is shorter than the eddy turnover time then the co-rotation torque saturates.

When the co-rotation torque saturates, the planet breaks free of the trap, and we compute its radial evolution due to the Lindblad torques along \citep{Seager10},\begin{align}
\Gamma &= C_\Gamma \Sigma_p \Omega_p^2 a^4 \left(\frac{M_p}{M_s}\right)^2\left(\frac{a}{|\Delta r|}\right)^2,
\label{eq:torq01}
\end{align}
where $\Omega_p$, $a$, and $M_p$ are the optical frequency, radius, and mass of the planet, $M_s$ is the mass of the host star, and $\Sigma_p$ is the gas surface density at the planet's location. The constant $C_\Gamma = -(2.34 -0.10 s)$ \citep{Paard10} has been determined through numerical simulations. The minimum length scale $|\Delta r|$ below which the gas does not contribute to Lindblad torques depends on the relative impact of the planet's gravity to the gas pressure on the dynamics of the gas. Below the Hill radius ($R_H$) gas particles become bound to the planet, while below the gas scale height ($H_p$) the gravitational effects of the gas are smoothed out. Therefore we set $|\Delta r| = max(H_p,R_H)$.

We evolve the orbital radius of the planet through $\dot{a}/a = \Gamma/J_p$ where $J_p = M_p a^2 \Omega_p$ is the orbital angular momentum of the planet. As such the planet's orbital radius evolves as, \begin{align}
\frac{1}{a}\frac{da}{dt} &= \frac{C_T \Sigma \Omega_p a^2}{M_p} \left(\frac{M_p}{M_s}\right)^2\left(\frac{a}{max(H_p,R_H)}\right)^2
\label{eq:torq02}
\end{align}
when the co-rotation torque saturates.

In what follows, when discussing the properties of our forming planets we will refer to the planet by the trap where it originated.

\ignore{
Generally the Lindblad torques cause the planet to migrate inward, while the co-rotation torque can cause the planet to migrate outward \citep{Paard14}. The net direction of the Type-I migration depends on the relative strength of the two torques, and hence the physical properties of the disk. At planet traps disk properties like the radial temperature profile or the dust opacity change such that the net direction of the angular momentum transport reverses - trapping planets at these transition points. 

The three planet traps that we model are the heat transition, water ice line and dead zone edge. These traps represent a transition in the temperature radial profile, dust opacity and turbulent parameter respectively. The location of the heat transition trap is defined in the analytic disk model as the disk radius at which the midplane temperature derived from viscous heating and direct irradiation are equal. The water ice line trap is defined as the disk radius where the water vapour abundance is equal to the water ice abundance. Finally the dead zone trap is defined as the radius where the Ohmic Elsasser number (see \cite{Crid16b}) is equal to 1. For further details see \cite{Crid16a}.
}

\subsection{ Importance of Opacity for Gas Accretion }\label{ssec:Opac}

The ability of a planet to accrete its atmosphere is limited by the cooling rate of the gas as it shreds its gravitational energy. This cooling rate depends on the gas opacity which measures the cross section of photon absorption or scatter per unit mass of gas. 

As the planet accretes gas it forms an envelope around the planet core that then slowly contracts onto the planet. This process is modeled by two physical properties: the gas accretion rate which depends on the planet mass and Kelvin-Helmholtz timescale, and the critical planet mass above which the gas envelope can contract onto the core. 

In previous population synthesis models (ie. \cite{HP13}) and in our previous work (CPA16) the Kelvin-Helmholtz timescale and critical mass are individually parameterized so that they have the form of Equation \ref{eq:mod02} and \begin{align}
M_{crit} = 10 \cdot f_{crit} \left(\frac{\mdot_{core}}{10^{-6} M_\oplus {\rm yr}^{-1}}\right)^q M_\oplus,
\label{eq:mod04}
\end{align}
respectively. \cite{Ikoma2000} showed that the parameters $c$ and $f_{crit}$ depend on the opacity, while the parameters $d$ and $q$ depend on the choice of opacity table \citep{IL04a}. 

The equivalent form of Equations \ref{eq:mod02} and \ref{eq:mod04} from \cite{Ikoma2000} have the form:\begin{align}
\tau_{env} = 3\times 10^8\left(\frac{M_{core}}{M_\oplus}\right)^{1-1/q}\left(\frac{\kappa}{{\rm cm}^2 {\rm g}^{-1}}\right)^{s/q} {\rm yr},
\label{eq:mod05}
\end{align}
where $s/q\simeq 1$, $\kappa$ is the dust opacity, and \begin{align}
M_{crit} = 7 \left(\frac{\mdot_{core}}{10^{-7} M_\oplus {\rm yr}^{-1}}\right)^q \left(\frac{\kappa}{{\rm cm}^2 {\rm g}^{-1}}\right)^s M_\oplus.
\label{eq:mod06}
\end{align}

The parameters $s$ and $q$ are within the range of 0.2-0.3 \citep{Ikoma2000}, and based on our choice of parameters for Equation \ref{eq:mod02} imply that $s=q=0.25$. In our normalization of Equation \ref{eq:mod02} we assume that $\kappa \sim 3.33 ~{\rm cm}^2 {\rm g}^{-1}$ which is consistent with the Rosseland mean opacity for a sub-mm grain computed by our opacity table.

With our assumed opacity we find that $f_{crit}\simeq 1.6$ while in CPA16 we assumed $f_{crit} = 0.2$ based on results from population synthesis models \citep{HP12}. When $f_{crit}$ is higher, gas accretion begins later in the planet's formation history which changes the radial location and timing of the atmosphere formation.

A smaller $f_{crit}$ assumes that the envelope opacity is lower, which would imply some grain growth within the envelope. Because the dust physics is not well constrained in the envelope of accreting planets we will show the results for $f_{crit} = 0.2$, which matches the parameters used in CPA16. In the Appendix we include the results for $f_{crit} = 1.6$ for the fiducial disk mass, connecting Equations \ref{eq:mod05} and \ref{eq:mod06} with an assumed constant envelope opacity of 3.33 cm$^2$g$^{-1}$.

\section{Results: Disk Astrochemistry}\label{sec:CDG}

The main purpose of this work is to assess the effect of different dust models on disk astrochemistry and the resulting atmospheric abundances. To that end we first compare the chemical results of the CPA16 disk model to the disk chemistry that is computed here, and based on the dust model presented in CPB17.

\subsection{ Comparison of the Distribution of Gas Species }

\begin{figure*}
\centering
\begin{overpic}[width=\textwidth]{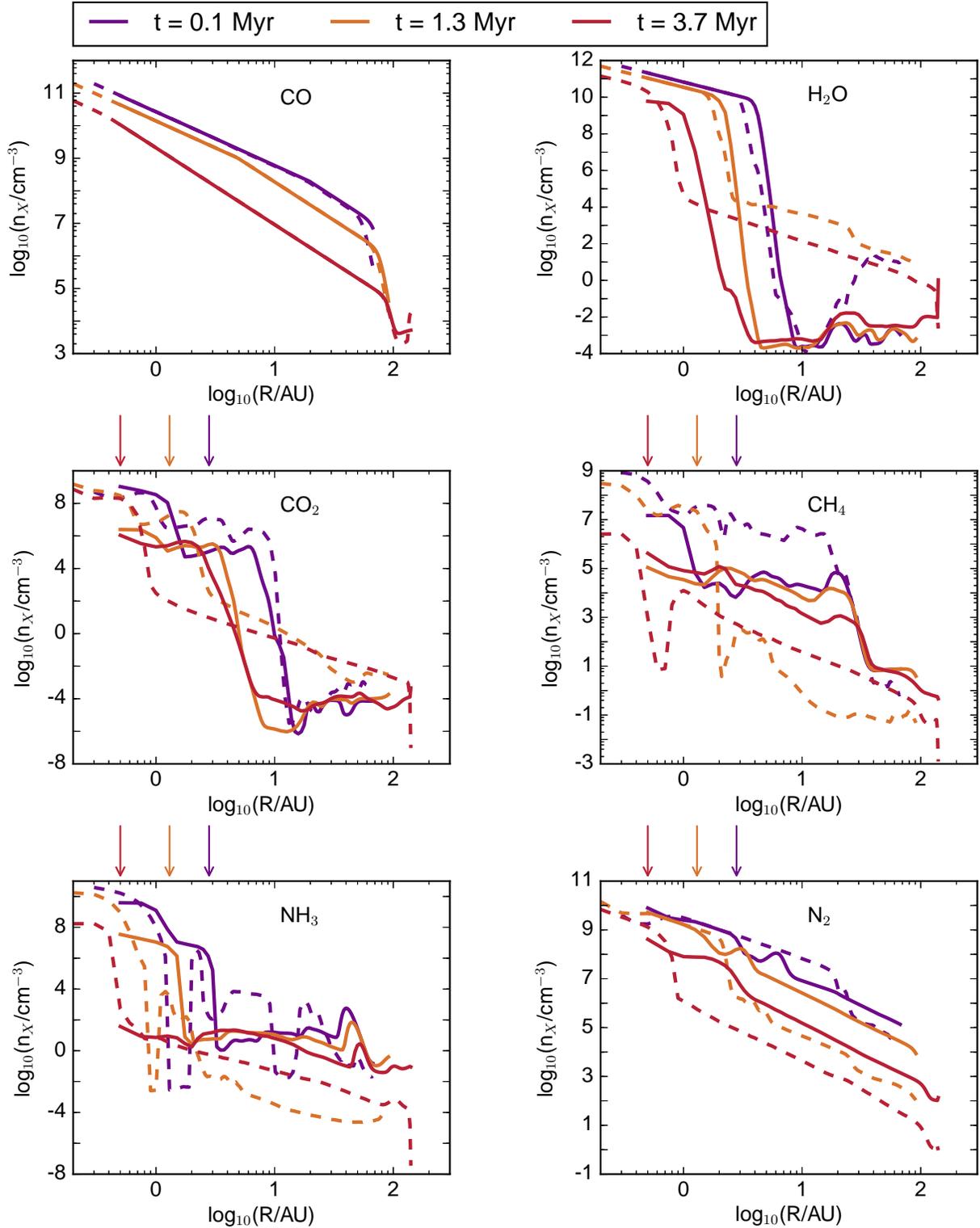}
\end{overpic}
\caption{ Radial distribution of the chemical species that are important to the observations of exoplanetary atmospheres for the CPA16 disk model (solid lines) and for the results of the CPB17 dust model presented in this work (dashed lines). The coloured arrows denote the approximate location of the water ice line in the CPB17 model, defined as the point where the water vapour abundance has dropped by half. }
\label{fig:chem01}
\end{figure*}

\begin{figure*}
\centering
\begin{overpic}[width=\textwidth]{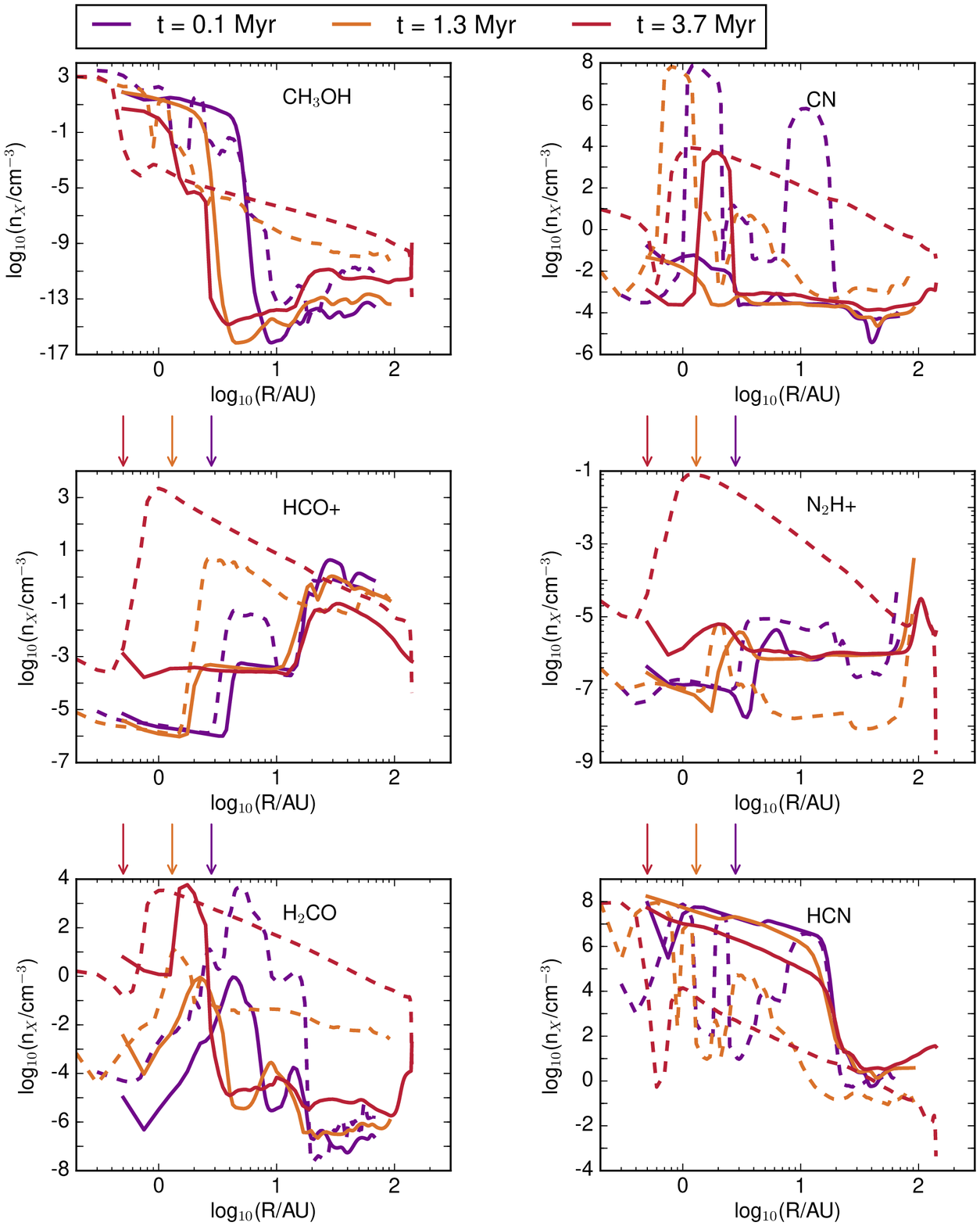}
\end{overpic}
\caption{ Same as in Figure \ref{fig:chem01} but for other chemical species that have been observed in protoplanetary disks, and are used as tracers of photochemical processes. }
\label{fig:chem02}
\end{figure*}

In Figures \ref{fig:chem01} and \ref{fig:chem02} we present the radial distribution of the midplane chemical abundances for a few chemical species in the CPA16 chemical model (solid line) and the chemical results from the disk model presented in CPB17 (dashed line). In Figure \ref{fig:chem01} we show the important chemical species to the observations of exoplanetary atmospheres, while in Figure \ref{fig:chem02} we show other molecules that have been observed or used as tracers in protoplanetary disks.

\subsubsection{ Primary Carbon and Oxygen Carriers }

In the top 4 panels of Figure \ref{fig:chem01} we compare the primary carbon and oxygen carriers in this disk midplane. An important difference between the chemical distributions that resulted from the CPA16 (solid line) and CPB17 (dashed line) dust models is the abundance of volatiles like H$_2$O and CO$_2$ at larger radii. For the results of this work, based on the dust model in CPB17, there is a higher abundance of volatiles farther out in the disk, where one would expect the ice component of the above species would dominate. The retention in the CPB17 dust model is connected to the fact that the largest dust grains are more rapidly depleted than in the CPA16 dust model. As a result the amount of dust available for volatile freeze out is reduced in the CPB17 dust model. In addition to the reduction of available surface onto which volatiles can freeze, a reduction of the disk opacity to UV photons can also lead to an enhancement of gas abundance through the photodesorption of frozen species.

At smaller radii than the water ice line there is an enhancement of dust surface density fueled by the inward radial drift of dust grains. This enhancement shifts the location of the water ice line inward by $\sim 0.5$ AU in the new dust model. We additionally find that the ice lines of CO and CO$_2$ are also shifted from their position in the CPA16 model. These changes are linked to the increase of photodesorption at larger radii than the water ice line that results from the reduced dust opacity in the outer disk, and higher radiative flux along the midplane.

\subsubsection{ Primary Nitrogen Carriers }

In the bottom 2 panels of Figure \ref{fig:chem01} we compare the abundances of N$_2$ and NH$_3$. As mentioned before (or see Figure \ref{fig:res04} below), there is a higher dust surface density within the water ice line of the CPB17 model when compared to the CPA16 model. This enhancement leads to higher abundance of nitrogen carriers within the water ice line because of the electron capture and subsequent dissociation of ions like N$_2$H$^+$ and NH$_4^+$ \citep{Herbst73} are catalyzed by the presence of dust. Similarly with a higher dust surface density at smaller radii, the gas is shielded from destructive high energy radiation. This dust enhancement is not seen in the CPA16 dust model because of the simpler treatment of dust physics and as a result we find that NH$_3$ is destroyed as the disk ages. 

An important caveat to our derived NH$_3$ abundances is that we currently ignore the formation of NH$_3$ ice through successive capture of atomic hydrogen by atomic nitrogen on the surface of grains (see for ex. \cite{Watson72}). This process is analogous to the formation of water ice on dust grains, which is included in our chemical model. The extra NH$_3$ ice could then contribute to its gas abundance through photodesorption and thermal desorption. This possible extra source of NH$_3$ formation implies that our presented abundances here may represent a lower bound of possible observed abundances.

\subsubsection{ Photochemical Tracers }

In the middle panels of Figure \ref{fig:chem02} we show the enhancement of chemical tracers of disk ionization (HCO$^+$ and N$_2$H$^+$) relative to the chemical model presented in CPA16. This enhancement confirms the high ionization rate that is present in the CPB17 disk model. In the bottom right panel of Figure \ref{fig:chem02} we show the abundance of HCN while in top right panel of Figure \ref{fig:chem02} we show the abundance of CN. Observationally, the relative abundance of these species has often been used as a proxy for UV flux. In this work we see a reduction in the abundance of HCN relative to the results of CPA16, and the corresponding enhancement of CN which suggests that UV photolysis is more prevalent in the CPB17 disk model than in our previous model. This prevalence is simply due to the reduced disk opacity and increased UV flux along the midplane. We will elaborate on the link between the flux of high energy and the abundances of HCN and CN in more detail in the following subsection.

At late times in the CPB17 disk model (red dotted line) tracers that are formed from photochemical processes (CN, HCO$^+$, N$_2$H$^+$, and H$_2$CO) follow similar radial distributions: low abundances within the water ice line and high abundance outside. At late times photochemistry dominates the productions of these chemical species, and hence their abundance is directly linked to the flux of high energy photons on the midplane. This fact can most easily be seen in HCO$^+$ which shows a dramatic increase in its abundance outside the water ice line, where high energy photons can first penetrate down to the midplane.  N$_2$H$^+$ is largely kept under-abundant by destructive reactions with gaseous CO \citep{Qi13,Qi15}. At the largest radii, outside of the CO ice line, N$_2$H$^+$ is recovered in all models when the CO gas freezes out.

\begin{figure}
\centering
\includegraphics[width=0.5\textwidth]{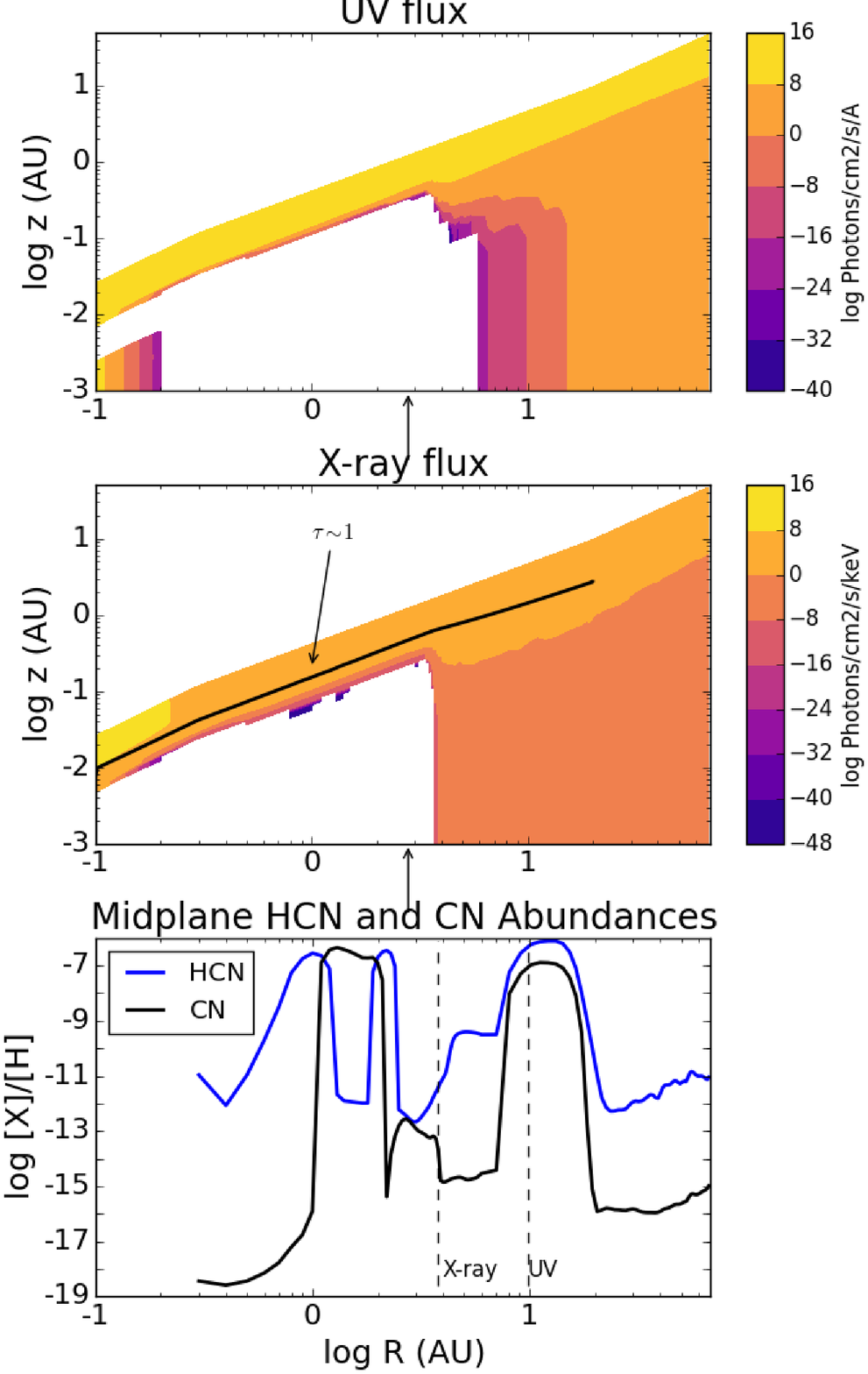}
\caption{The radiative flux of the highest energy bands of UV (top panel) and X-ray (middle panel) along with the midplane abundances of HCN and CN at $\log_{10} t/yr = 5.0$. On the middle panel, we plot an estimate of the height where an incoming X-ray photon reaches an optical depth of one. On the bottom panel, the dotted lines show the midplane location of the $10^{-8}$ contours of the UV and X-ray flux. The location of the water ice line is denoted with the black arrow on the top and middle panels. On the contour plots, regions of white space denote zero flux.}
\label{fig:radfield}
\end{figure}

\subsection{ HCN/CN as a Tracer of Dust Physics and Radiative Flux }

As an example of the subtle link between dust physics, radiative transfer, and chemistry we discuss the changes in the abundance of HCN and CN at different disk radii. The relative abundance of HCN and CN is often used as an observational tracer of UV flux in protoplanetary disks (for ex. \cite{Kastner1997}) because of the chemical pathway that produces CN through the photodissocation of HCN \citep{WL2000}.

In Figure \ref{fig:radfield} we relate the midplane abundance of HCN and CN to the flux of high energy photons through the disk. Within the water ice line (denoted by the black arrows) the large abundance of dust grains shields the gas from high energy photons allowing ion chemistry, catalyzed by the presence of dust, to dominate the production and destruction of HCN and CN. The ions responsible for these reactions are the residual ions from our initial chemical state (Table \ref{tab:initchem}). Crossing the water ice line results in a strong reduction of dust surface density, and a subsequent reduction in the abundances of both HCN and CN.

The radiative flux of both UV and X-ray photons show similar features in our disk models. There is a layer of highly ionized  (electron fraction $\sim 10^{-2}$) gas far above the midplane of the disk. This layer is denoted by the brightest contours in Figure \ref{fig:radfield}. In the middle panel of Figure \ref{fig:radfield} we plot an estimate of the height where photons reach an optical depth of one. Lower than this height the flux of photons coming directly from the source (ie. ignoring scattering) is attenuated, reaching zero in the white regions of the figure. The high optical depth of the disk's dust within the water ice line creates a shadowing outside the ice line which is eventually filled in through scattering. As the disk ages and the dust surface density is reduced, the shadowing shrinks and its edge moves to lower radii. X-ray photons scatter most efficiently, so their flux recovers very quickly just outside the location of the water ice line. By contrast, the lower energy UV 
photons take longer to recover along the midplane. We note the difference in the radial location of the $10^{-8}$ contour as dotted lines on the bottom panel of Figure \ref{fig:radfield}.

Moving outward from the water ice line, the abundance of HCN and CN initially drops because of the reduced surface density of dust, but HCN begins to recover when the flux of X-rays along the midplane suddenly increases. The abundance of CN remains low until the flux of UV photons begins to rise, at which point it is produced by the photodissociation of HCN. While HCN is destroyed by photodissociation, it is efficiently produced through ion chemistry, maintaining its abundance until the gas is sufficiently cold to allow HCN to freeze out. Finally at the largest radii we reach the HCN ice line, where it begins to freeze out onto dust grains. This freeze out also strongly reduces the abundance of CN since there is less HCN in the gas phase to photodissociate.

\subsection{ Summary of Key Chemical Results }

\begin{itemize}
\item We find minimal changes to the location of the water and CO ice line between the two disk models from CPA16 and CPB17
\item The CO$_2$ ice line shows larger variation between disk models, caused by the increased flux of UV at the midplane
\item Volatiles are more abundant in the outer disk of the CPB17 disk model, caused by a radial drift induced reduction of dust surface density in the outer disk and less efficient freeze out
\item An enhancement of NH$_3$ and N$_2$ is found within the water ice line of the CPB17 disk model because of the higher dust surface density within the water ice line
\item The abundance of photochemical tracers (HCO$^+$, HCN, and CN) are indicative of higher UV and X-ray flux on the midplane of the CPB17 disk model, outside of the water ice line
\item Strong variation in chemical abundances can be linked to gradients in dust surface density, and radiative flux of high energy photons
\end{itemize}

\begin{figure*}
\centering
\subfigure[ Planetary formation tracks from CPA16. ]{
	\label{fig:res02a}
	\includegraphics[width=0.5\textwidth]{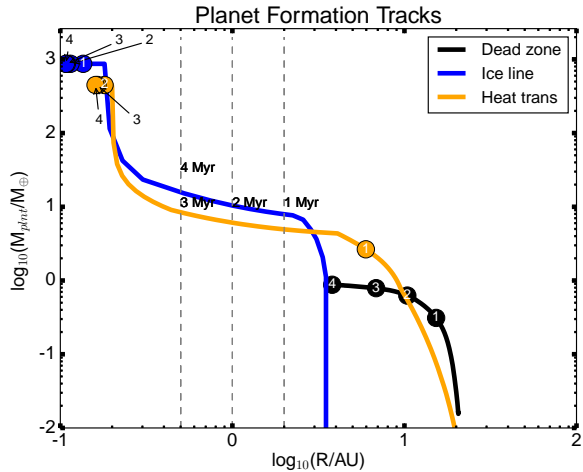}
}
\subfigure[ Planetary formation tracks for this work based on the CPB17 dust model. ]{
	\label{fig:res02b}
	\includegraphics[width=0.5\textwidth]{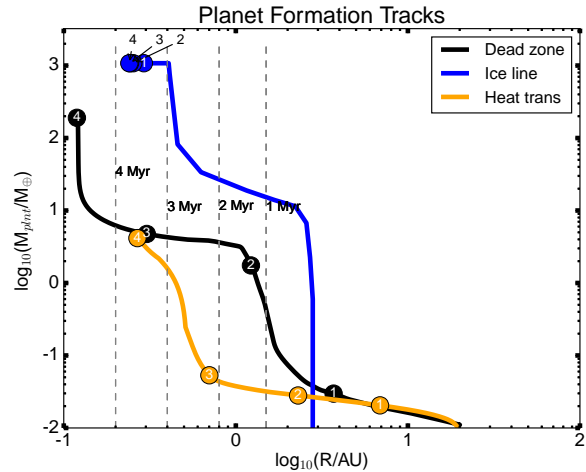}
}
\label{fig:res02}
\caption{ Planet formation tracks for the two different disk models presented in CPA16 and CPB17. Each model had the same disk mass, size, gas evolution and planet formation model. The only difference is in the choice of dust model. The grey dashed lines shows the location of the water ice line for the two disk models at 1,2,3 and 4 Myr. The annotated numbers shows where on the mass and semi-major axis diagram the planet is at 1,2,3 and 4 Myr. In the CPB17 dust model, the chemical abundance of the disk was computed at a higher resolution than in CPA16. Hence why the water ice line appears to evolve to smaller radii in the right panel. }
\end{figure*}

\begin{figure*}
\centering
\subfigure[Atmospheric results presented in CPA16.]{
	\label{fig:res01a}
	\includegraphics[width=0.5\textwidth]{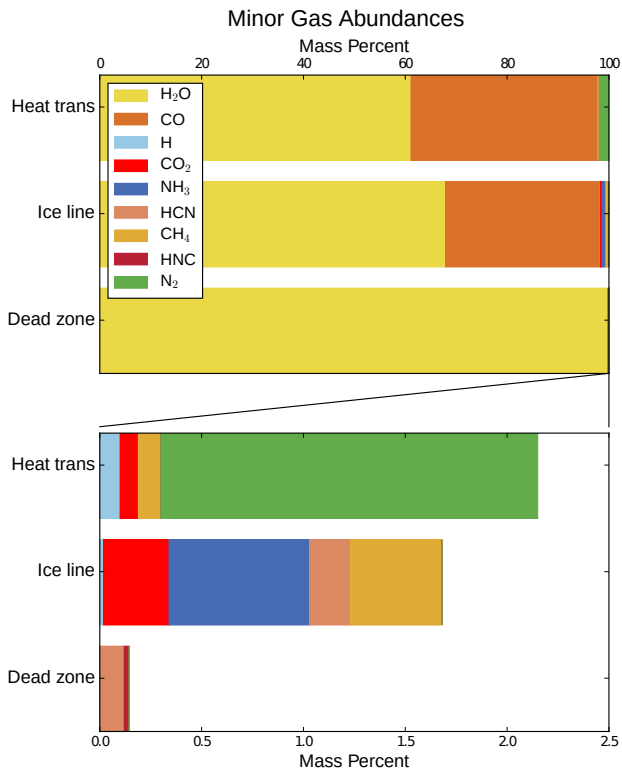}
}
\subfigure[Atmospheric results of this work, based on the dust model presented in CPB17.]{
	\label{fig:res01b}
	\includegraphics[width=0.5\textwidth]{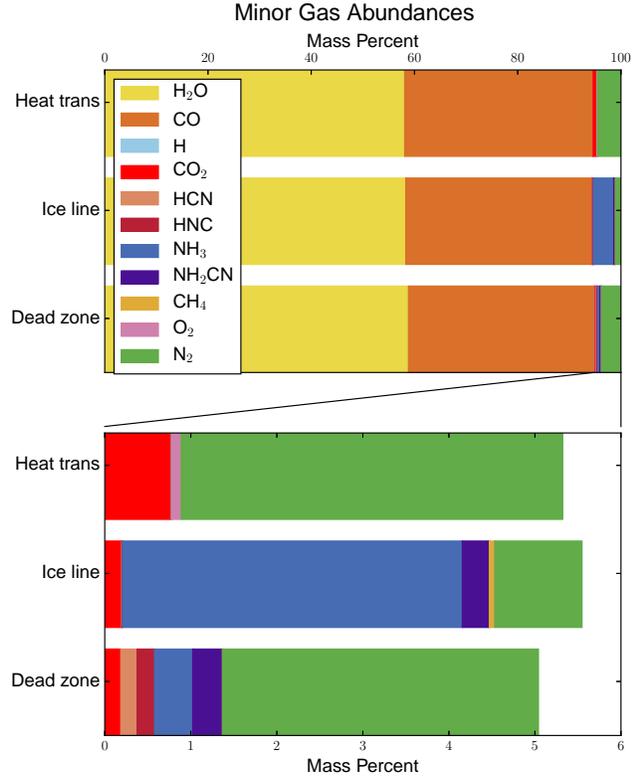}
}
\label{fig:res01}
\caption{ Atmospheric abundances for most abundant gases other than hydrogen and helium. The bottom panels of both plots are the most abundant gases after water and CO are also removed. Note that in the bottom panel of Figure \ref{fig:res01b} the scale is a factor of 5 larger than in the bottom panel of Figure \ref{fig:res01a}. }
\end{figure*}

\begin{table}
\begin{center}
\caption{Final planetary properties at $t=t_{life}$ from the CPA16 and CPB17 dust models}
\begin{tabular}{l c c c}\hline
{\bf CPA16} & $M_{plnt} ~(M_\oplus)$ & $a_{final}~({\rm AU})$ \\\hline
Dead zone & $ 0.95 $ & $3.7$ \\\hline
Ice line & $ 753.2 $ &  $0.11$ \\\hline
Heat transition & $ 454.5 $ & $0.15$ \\\hline
{\bf CPB17} & $M_{plnt} ~(M_\oplus)$ & $a_{final}~({\rm AU})$ \\\hline
Dead zone & $ 197.2 $ & $ 0.12 $ \\\hline
Ice line & $ 1090.2 $ & $ 0.25 $ \\\hline
Heat transition & $ 4.24 $ & $ 0.27 $ \\\hline
\end{tabular}
\label{tab:res01}
\end{center}
\end{table}

\section{ Results: Planetary Atmospheres }\label{sec:res01}

In the CPB17 disk model, the evolution of the chemical state of the disk changes the formation history of dynamically trapped planets. Here we compare the planetary atmosphere results from CPA16 model and the chemical model presented in this work.

\subsection{ Comparing Planetary Formation and Atmospheric Composition for Different Dust Models }

In Figures \ref{fig:res02a} and \ref{fig:res02b} we show the evolution of the forming planets through the mass and semi-major axis diagram for the model presented in CPA16 (\ref{fig:res02a}) and the model presented in CPB17 (\ref{fig:res02b}). We find a significant difference in the planet formation histories between these two disk models.

At the water ice line, the dust density is enhanced in the new dust model. This enhancement can increase the dust-to-gas ratio up to 1-to-4 from the standard 1-to-100 that is used in CPA16, which drastically increases the rate of solid accretion in the new dust model. Because of the enhancement the planet forming at the water ice line builds a larger core and rapidly draws down an atmosphere, faster than in the CPA16 model. Formation happening quicker implies that the planet will not migrate inwards as far in the CPB17 dust model which results in the larger final semi-major axis for the planet in the water ice line. 

The planet forming at the dead zone edge in the CPA16 model did not accrete an atmosphere because it spent its whole formation history far out in the disk. This is in contrast to the CPB17 model where the dead zone edge rapidly evolves inward, bringing its forming planet with it. In the CPB17 model the dead zone planet results in a mass of approximately 2 Saturn masses at $\sim 0.12$ AU. 

The heat transition planet in the CPB17 model results in a super Earth sized planet very close to its host star. The mass of this planet is significantly smaller than its counterpart in the CPA16 model planet which is approximately 1.4 Jupiter masses. The difference between these two models is due to the evolution of the dead zone edge which crosses the position of the heat transition trap very early in the CPB17 model. In this model, the dead zone spends most of its time at smaller radii than the orbital radius of the heat transition trap. As a result the planet remains trapped at the heat transition trap location for its entire formation history, staying at larger radii than its counterpart in the CPA16 model. At larger radii, the surface density of gas and dust is lower and hence the planet has less material on which to feed.

In Figure \ref{fig:res01a} we show the resulting chemical abundances of all minor gases (most abundant gases other than H$_2$ and He) produced in the CPA16 model. Comparatively in Figure \ref{fig:res01b} we show the resulting chemical abundances for the planets formed in the CPB17 model. Since the planet at the dead zone edge did not accrete an atmosphere in the CPA16 model, we instead plot the abundance of ice species that were accreted and assumed 100$\%$ out-gassing. We do not believe this is an appropriate assumption for the production of an atmosphere around this planet and hence will limit our discussion to planets that directly accreted their atmosphere from the disk gas. 

In our new dust model the water and CO abundances are similar because all planets accreted their atmospheric gas near or within the water ice line. As the disk ages and cools the three planet traps converge at small radii and since the heat transition and dead zone planets accrete their gas later in the disk lifetime, all planets accrete gas with similar chemical abundances.

The less dominant species show some variation between the two dust models. The heat transition planet accretes its gas at a very late time at a colder region of the disk. This results in a negligible amounts of methane and elemental hydrogen, but a small quantity of oxygen gas being accreted. The ice line planet accreted its atmosphere in less than 1 Myr in both models, however it accreted its solid core faster in the CPB17 model, and did not migrate as close to the host star as it did in the CPA16 model. Because of the slower migration the CPB17 ice line planet accreted its gas in a slightly cooler position of the disk compared to its CPA16 counterpart which resulted in less methane, but more nitrogen gas being accreted. 

\begin{table}
\begin{center}
\caption{ Carbon-to-Oxygen (C/O) and Carbon-to-Nitrogen (C/N) ratios for the planets from CPA16 and this work based on the dust model from CPB17. }
\begin{tabular}{ l c c }
\hline
{\bf CPA16} & C/O & C/N \\\hline
Dead zone & N/A & N/A \\\hline
Ice line & 0.23 & 47 \\\hline
Heat transition & 0.28 & 10 \\\hline
{\bf THIS PAPER} & C/O & C/N \\\hline
Dead zone & 0.29 & 4.12 \\\hline
Ice line & 0.29 & 4.08 \\\hline
Heat transition & 0.29 & 4.17 \\\hline
\end{tabular}
\label{tab:res02}
\end{center}
\end{table}

\subsubsection{ C/O and C/N Elemental Ratios }

Two important tracers of planet formation history are shown in Table \ref{tab:res02}. The elemental ratios of carbon, nitrogen and oxygen trace {\it when} and {\it where} the planetary atmosphere was accreted. The carbon-to-oxygen ratio (C/O) indicates whether a planet accretes its atmosphere at smaller orbital radii than the water ice line, or a larger radii. Both the CPA16 and CPB17 models result in planets that accrete their atmosphere close to the water ice line in the disk, and hence they all have similar C/O which is similar to the disk's C/O (0.288). This sub-solar C/O results from our choice of initial chemical abundances which we chose to replicate the chemical state of molecular clouds (ie. \cite{AH99}). In doing so we assume that the chemical abundance of the gas is not reset during the disk formation process - an effect that can significantly alter the chemical state of the disk (see for ex. \cite{Eistrup2016}).

The carbon-to-nitrogen ratio is also sensitive to the location of the planet when it accreted its atmosphere relative to the ice lines of nitrogen volatiles. While the abundance of nitrogen carriers (NH$_3$, HCN, N$_2$) is dependent on when the planet accretes its gas, C/N is more dependent on our choice of disk model. In the CPB17 disk model, more nitrogen (by mass) is accreted which reduces C/N by a factor of between 2.5 and 10 relative to the results of the CPA16 model. 

\begin{figure*}
\centering
\subfigure[Planetary formation tracks for $M_{disk,0} = 0.08 M_\oplus$]{
	\label{fig:res03a}
	\includegraphics[width=0.5\textwidth]{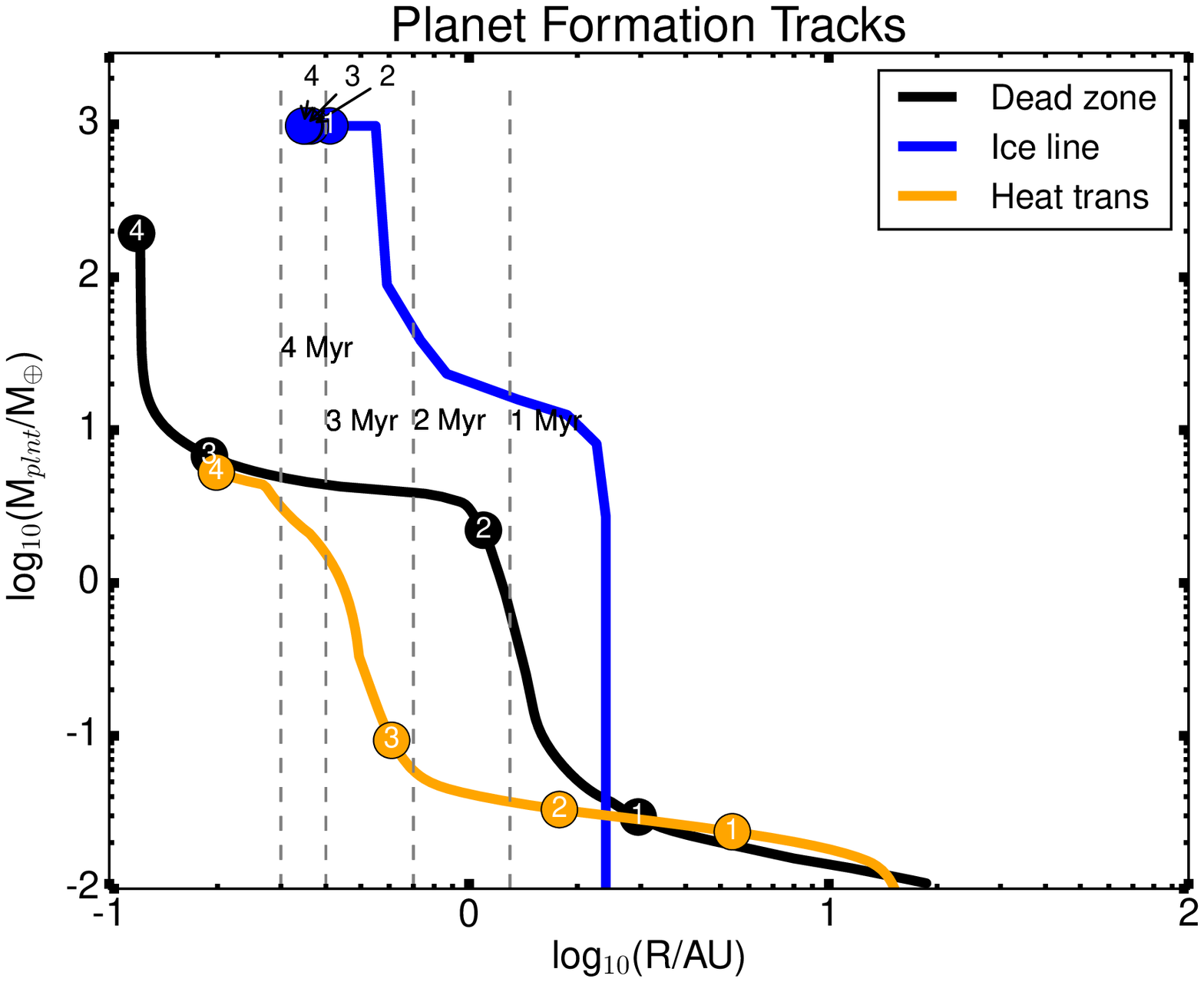}
}
\subfigure[Planetary formation tracks for $M_{disk,0} = 0.12 M_\oplus$]{
	\label{fig:res03b}
	\includegraphics[width=0.5\textwidth]{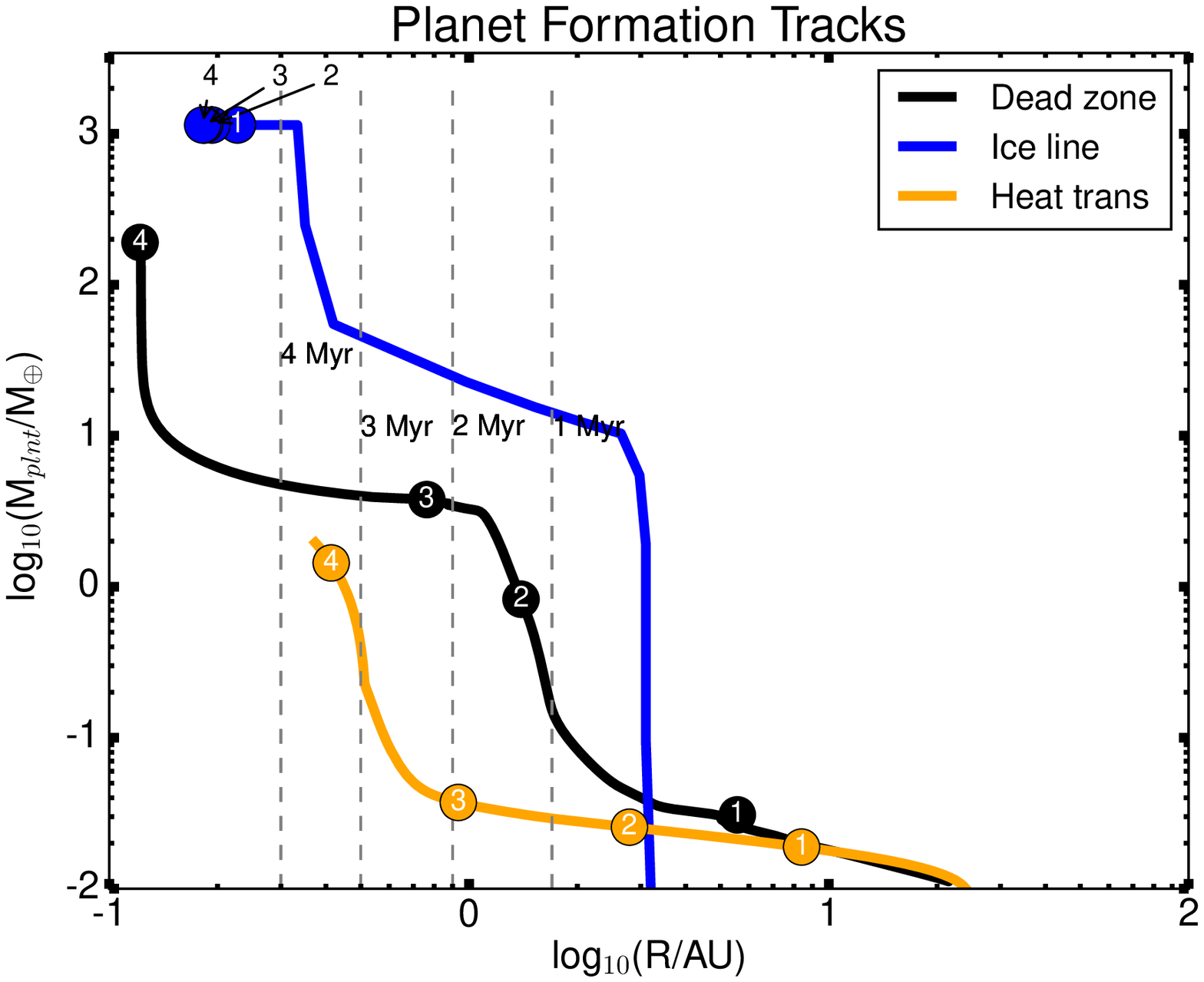}
}
\label{fig:res03}
\caption{ Planet formation tracks for the two additional disk models (see \S \ref{sec:res02}), annotated in the same as in Figures \ref{fig:res02a} and \ref{fig:res02b}. }
\end{figure*}

\subsection{ Varying Initial Disk Mass }\label{sec:res02}

\begin{table}
\begin{center}
\caption{ Final planetary parameters for the models of higher and lower initial disk masses. }
\begin{tabular}{l c c}\hline
$M_{disk,0} = 0.08M_\oplus$ & $M_{plnt} ~(M_\oplus)$ & $a_{final}~({\rm AU})$ \\\hline
Dead zone & $197.4$ & $0.12$ \\\hline
Ice line & $1095.0$ & $0.98$ \\\hline
Heat transition & $6.0$ & $0.19$ \\\hline  
$M_{disk,0} = 0.12M_\oplus$ & $M_{plnt} ~(M_\oplus)$ & $a_{final}~({\rm AU})$ \\\hline
Dead zone & $203.5$ & $0.13$ \\\hline
Ice line & $1194.0$ & $0.23$ \\\hline
Heat transition & $1.9$ & $0.37$ \\\hline  
\end{tabular}
\label{tab:res03}
\end{center}
\end{table}

As a prelude to a full population synthesis model, we present the results of two additional disk models with a higher (0.12 M$_\odot$) a lower (0.08 M$_\odot$) initial disk mass than our fiducial model (0.1 M$_\odot$).

\ignore{
An important feature of planet formation theories is exploring a range of disk parameters and their effect on the observable populations of exoplanets. This process is known as population synthesis which focuses on recreating the distribution of planets in the mass semi-major axis diagram that is produced by observations.  Population synthesis is also a useful tool for understanding what range of chemical abundances are expected in our disk and formation model.

A full population synthesis model would possibly include $>$ 1000 individual disk models which is too computationally expensive for our current method (one chemical model requires $\sim$ 3-4 weeks to run at the required temporal resolution). As a first step we ran two disk models, one with a higher initial mass (0.12 M$_\odot$) and the other with a lower initial mass (0.08 M$_\odot$) than our fiducial model (0.1 M$_\odot$). 
}

The initial disk mass sets the normalization for the initial gas surface density profile for a given initial disk size. Because of its dependence on gas surface density, the initial mass accretion rate is also set by the choice of initial disk mass. In our disk model, the temperature of the inner (r $\lesssim 20$ AU) disk is dependent on viscous heating, and hence is also dependent on the initial disk mass. All of these dependencies conspire to change the initial location of each of the planet traps, and the formation history for the planets forming within the traps.

In Figures \ref{fig:res03a} and \ref{fig:res03b} we show the planet formation tracks for the disk models with initial mass $0.08 M_\oplus$ and $0.12 M_\oplus$ respectively. In the lower mass disk, the heat transition and dead zone planets begin at smaller orbital radii than the planets in the high mass disk. As a result the planets in the lower mass disk accrete their material in a region of the disk with higher gas and dust surface density, building the core of the planet faster. In the case of the planet trapped at the heat transition trap, because its core is built faster in the lower mass disk it has time to accrete an atmosphere. While in the higher mass disk model the heat transition planet did not. The ice line planet also accretes its core faster in the low mass disk model, the reason for this is discussed below.

The final masses and semi-major axes are shown in Table \ref{tab:res03}. The planet forming at the dead zone trap accretes a similar mass and ends its evolution at a similar semi-major axis in all three models. This feature is related to the strength of the dust surface density enhancement that occurs at the ice line, and the tendency of the dead zone to evolve towards this enhancement. This tendency strongly ties the evolution of both traps at the latest times in the disk, and since the planet in the dead zone trap never accretes enough material to open a gap, it is bound to the radial evolution of its trap. 

At late times the viscous regime shrinks and eventually disappears completely, resulting in a disk that is heated solely by direct irradiation from its host star. The resulting temperature structure does not evolve in time and hence the location of the ice line and dead zone evolve to nearly the same radius in every disk model before ceasing its radial evolution.

\ignore{
\begin{figure}
\centering
\includegraphics[width=0.5\textwidth]{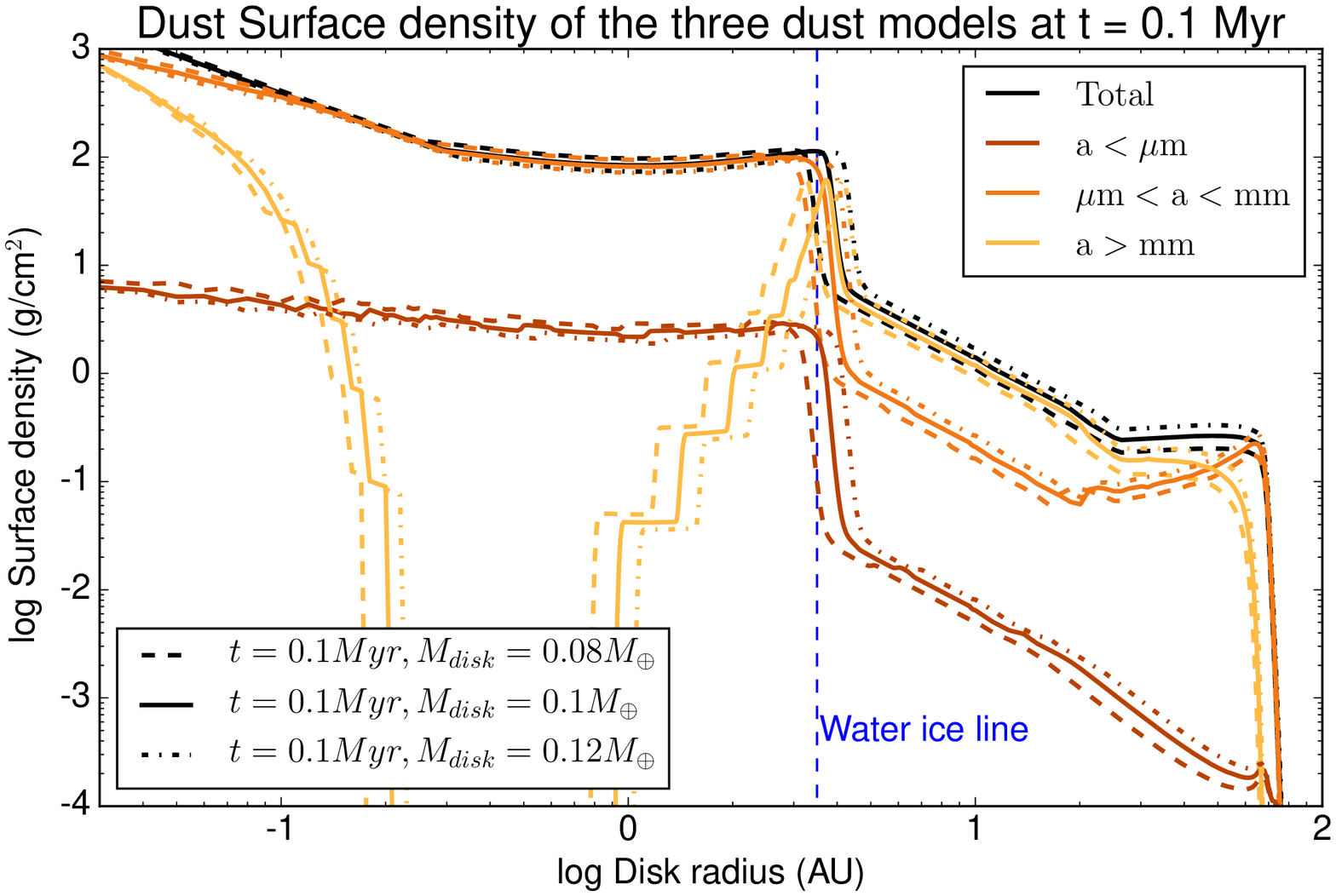}
\caption{ Dust surface density for the three presented models. Along with the total surface density (in black) the densities are binned into separate sizes: sub-micron, sub-millimeter and sizes greater than a millimeter. }
\label{fig:res04}
\end{figure}
}

\begin{figure*}
\centering
\subfigure[]{
	\label{fig:res04a}
	\includegraphics[width=0.5\textwidth]{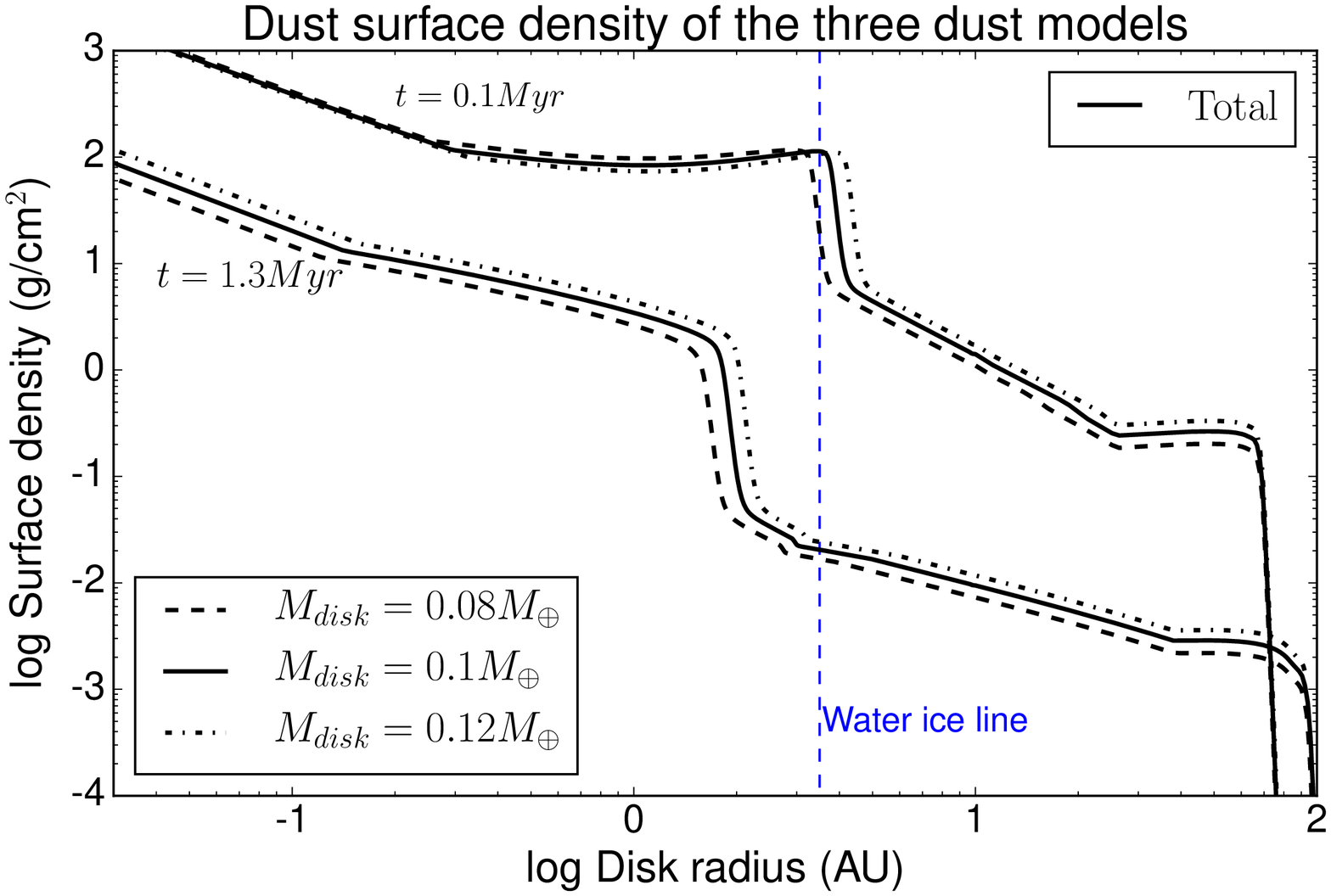}
}
\subfigure[]{
	\label{fig:res04b}
	\includegraphics[width=0.5\textwidth]{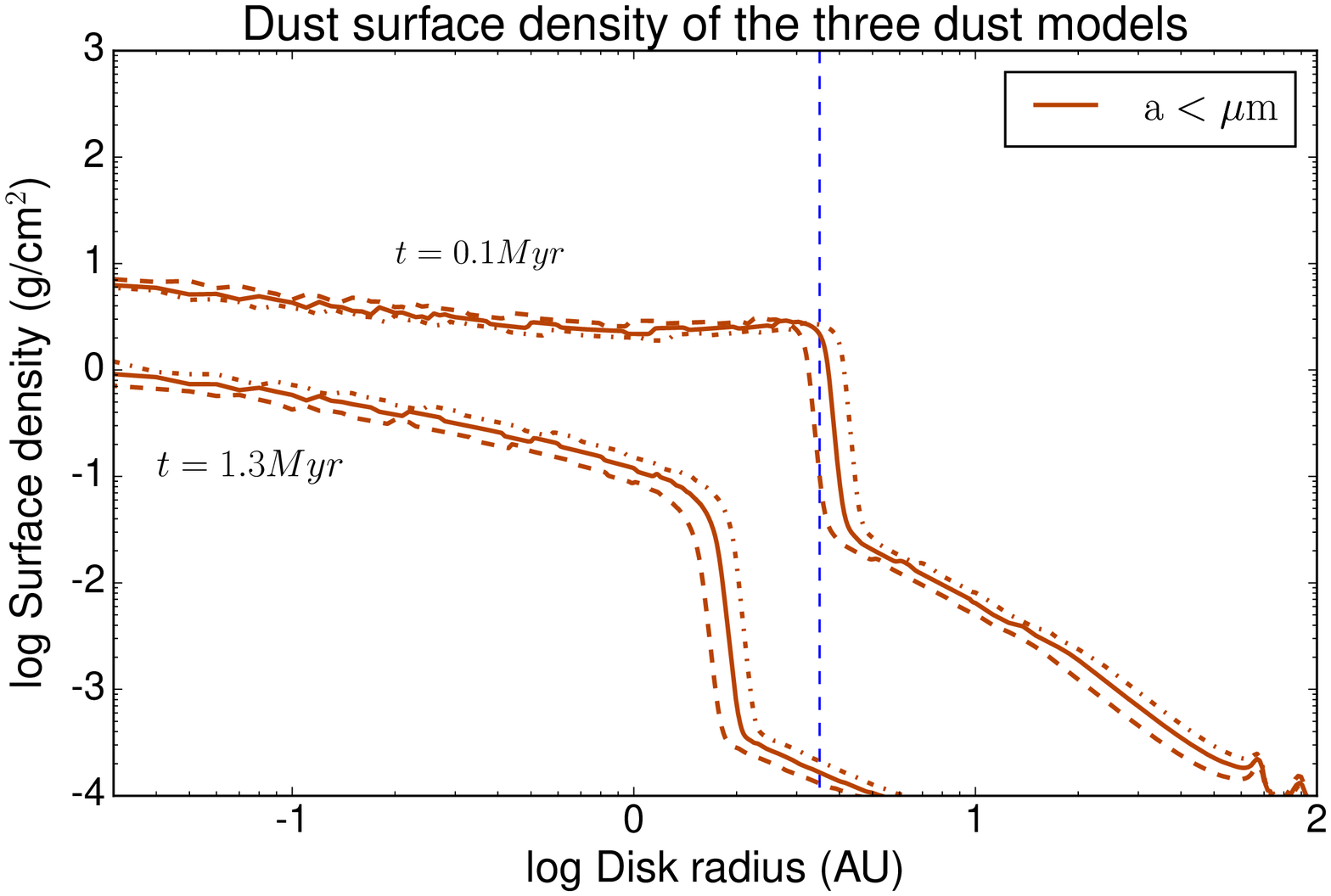}
}
\subfigure[]{
	\label{fig:res04c}
	\includegraphics[width=0.5\textwidth]{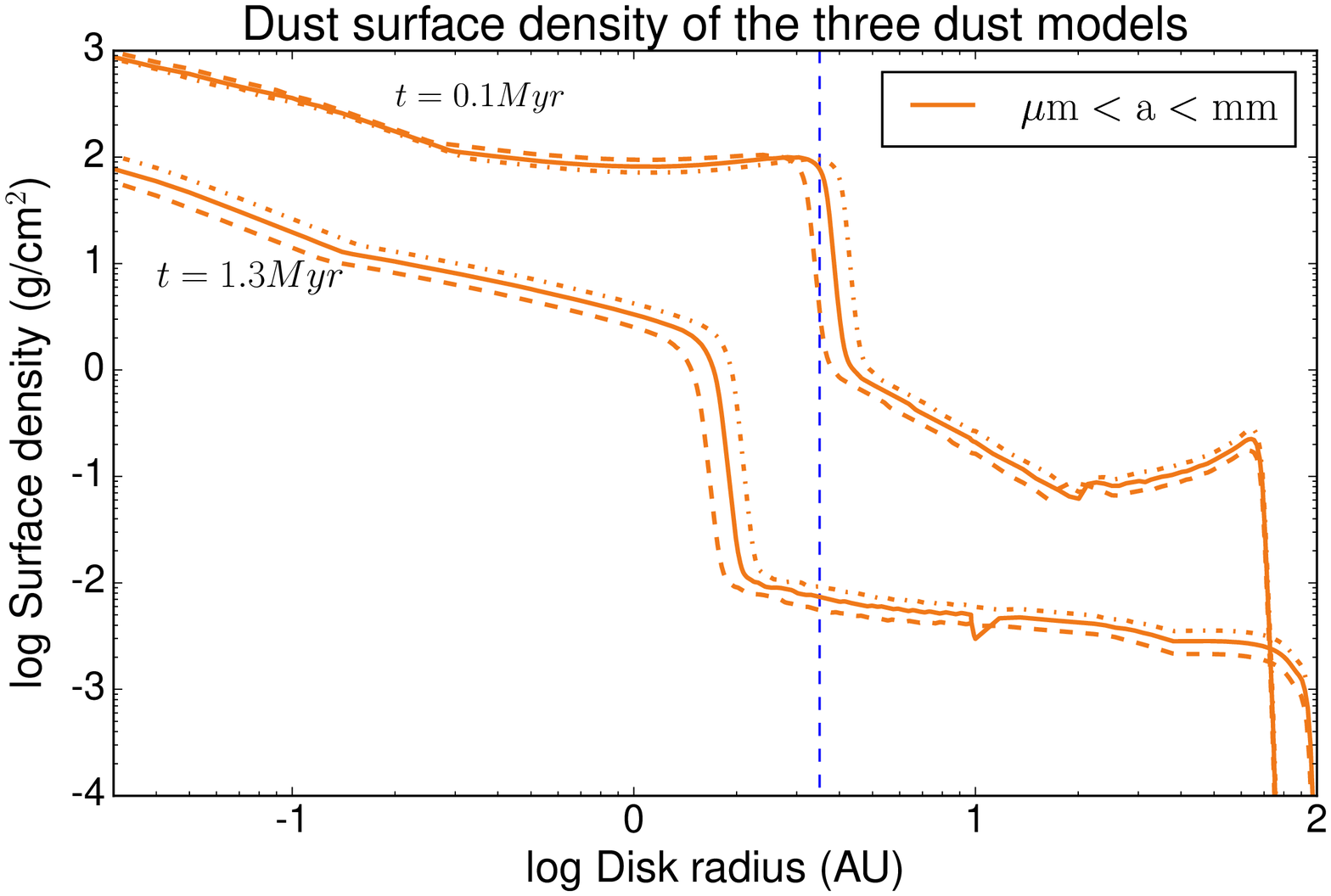}
}
\subfigure[]{
	\label{fig:res04d}
	\includegraphics[width=0.5\textwidth]{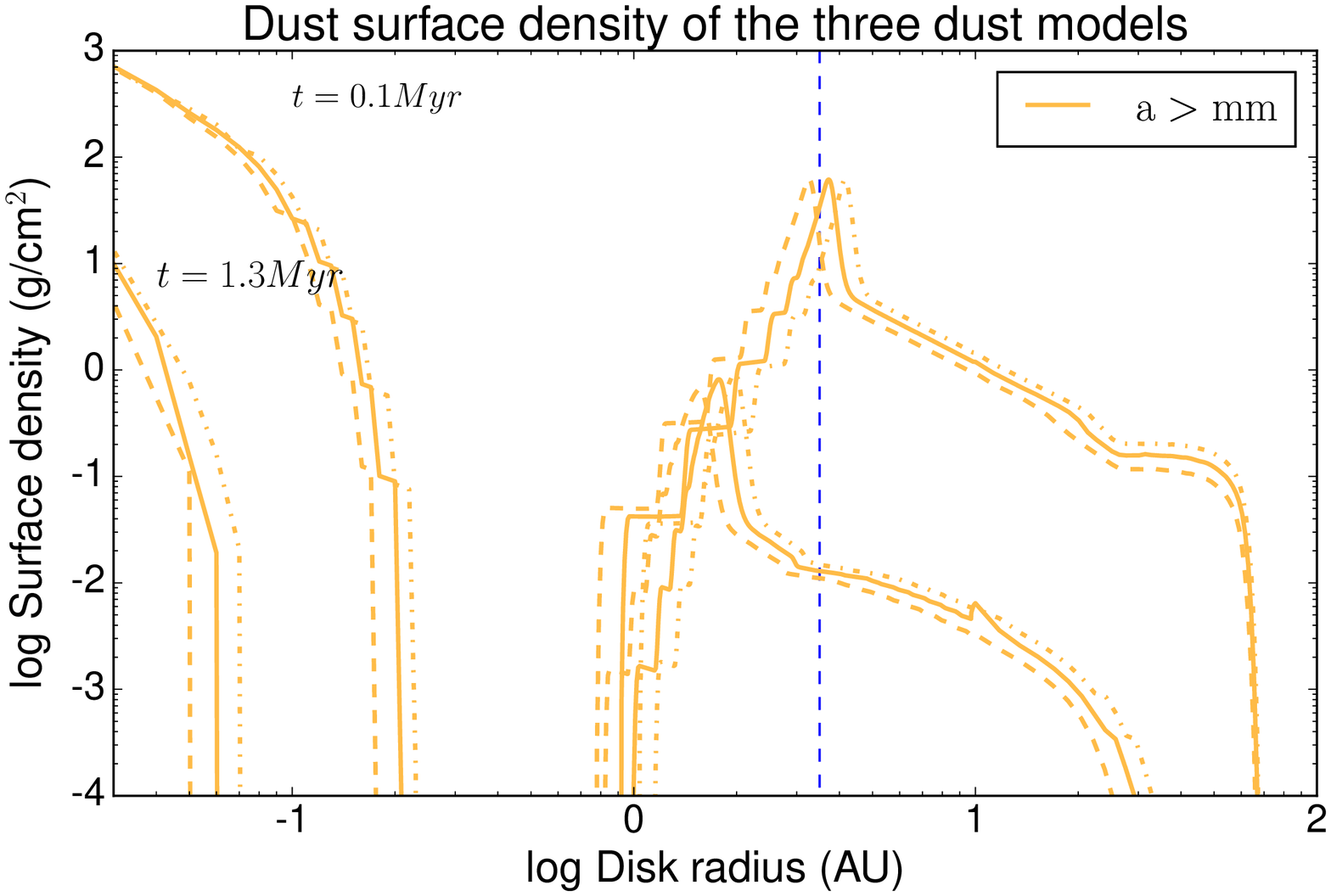}
}
\caption{ Dust surface density for the three presented models. Along with the total surface density (in black) the densities are binned into separate sizes: sub-micron, sub-millimeter and sizes greater than a millimeter. These dust surface densities are shown for two snapshots of our disk model, with 0.1 Myr represented by the top line in all figures, and 1.3 Myr represented by the bottom lines. Dust is cleared  between the two snapshots because of radial drift and viscous evolution. This clearing is particularly effective on the largest grains (Figure \ref{fig:res04d}), which shows a truncation in the disk at larger radii. }
\label{fig:res04}
\end{figure*}

\subsubsection{ Variation in Dust Surface Density Distribution }

In Figure \ref{fig:res04} we show the dust surface density for the three disk models shown at 0.1 Myr and 1.3 Myr. In the figures we denote the location of the water ice line for the fiducial disk model. The water ice line for the two other models is initially located at smaller orbital radii for the lower mass disk and a larger radii in the higher mass disk. In the outer disk ($r > r_{iceline}$) the dust surface density scales with the total initial disk mass with higher initial disk masses leading to higher dust surface densities, while in the inner disk ($r < r_{iceline}$) the reverse is true. 

This inversion is related to the location of the ice line in each of the disk models. The radial dependence of the dust surface density is dependent on the location of the water ice line, because of its impact on the fragmentation threshold speed of the grains. Hence the radial offset between dust models is partly due to the radial offset of the water ice line. On top of the radial offset, radial drift is more efficient in the lower mass disk because of the reduced gas surface density. 

The radial drift speed scales roughly linearly with the Stokes number (St, when St $< 1$), which describes the aerodynamically properties of the dust grain and hence the coupling to the gas. The dependency of the Stokes number on disk surface density and temperature is contingent on which physical process sets the maximum grain size. When the radial drift dominates the maximum size of the grains (ie. grains drift faster than they fragment) the Stokes number is inversely proportional to the gas surface density. However, when fragmentation dominates, the Stokes number depends only on the gas temperature \citep{B12}. Hence at large radii where the radial drift dominates, the lower gas surface density of the low initial mass disk results in larger Stokes numbers and more efficient radial drift. This faster radial drift brings more material to the inner disk, producing higher dust surface densities in the inner disk than the higher initial disk mass models.

The higher dust surface density in the low mass disk is what leads to the faster evolution of the ice line planet shown in Figure \ref{fig:res03a}. The planet starts its gas accretion with a heavier core and hence undergoes a shorter period of slow gas accretion. Because this slow gas accretion phase is responsible for the majority of semi-major axis evolution for forming planets, the resulting gas giant does not evolve far from its initial position.

Also shown in the figure is the temporal evolution of the dust surface density distributions from the earliest time (0.1 Myr) to an intermediate time (1.3 Myr) through the lifetime of the disk. As the disk ages, radial drift and viscous evolution causes the surface density of the dust to drop at all radii. Interestingly, the outer edge of the large dust grain disk (see Figure \ref{fig:res04d}) shrinks, while the outer edge of the intermediate sized grains ($\mu m < a < mm$, see Figure \ref{fig:res04c}) grows. 

\subsubsection{ Resulting Atmospheric Chemical Abundances }

\begin{figure*}
\centering
\subfigure[Final bulk composition for $M_{disk,0} = 0.08 M_\oplus$]{
	\label{fig:res05a}
	\includegraphics[width=0.5\textwidth]{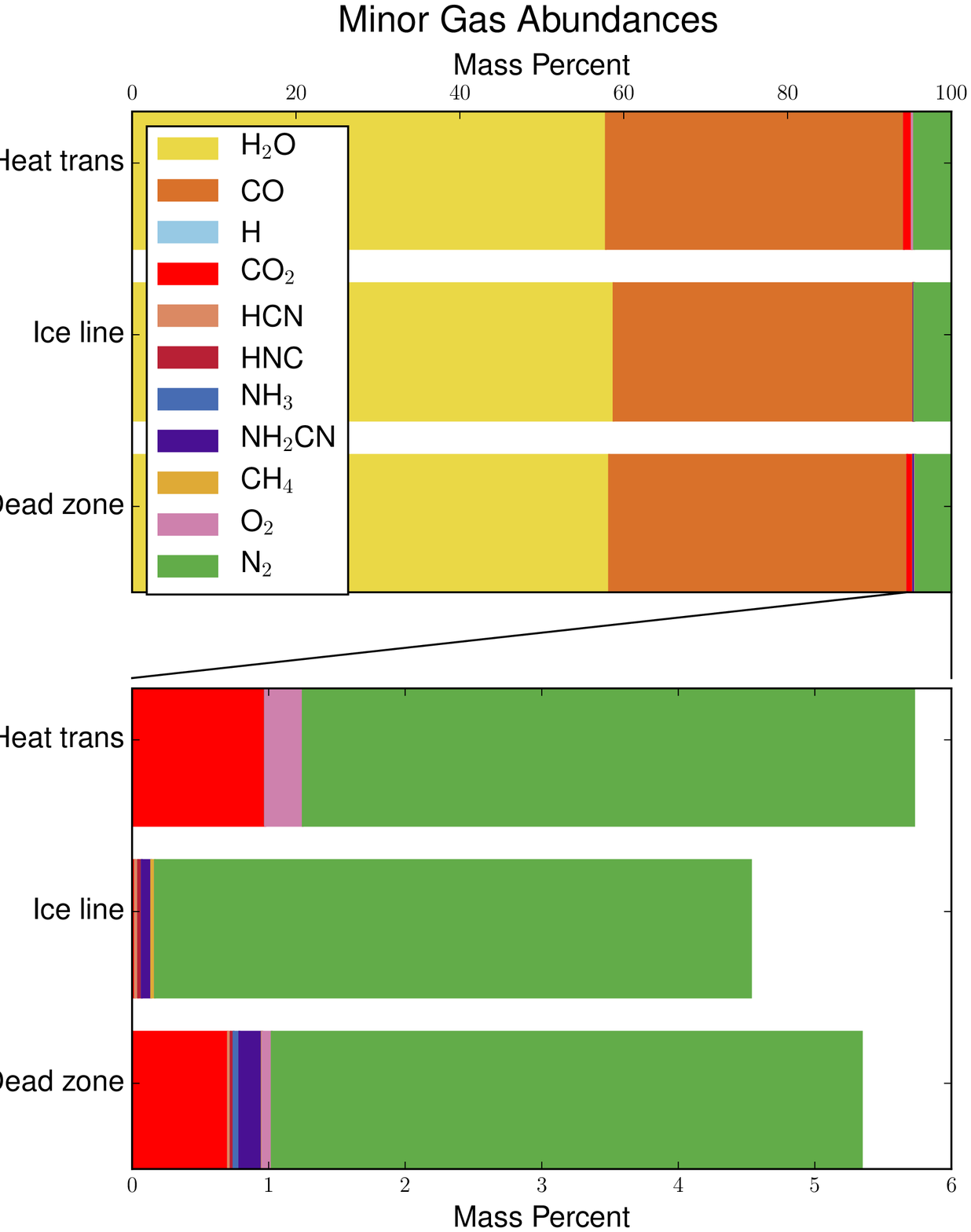}
}
\subfigure[Final bulk composition for $M_{disk,0} = 0.12 M_\oplus$]{
	\label{fig:res05b}
	\includegraphics[width=0.5\textwidth]{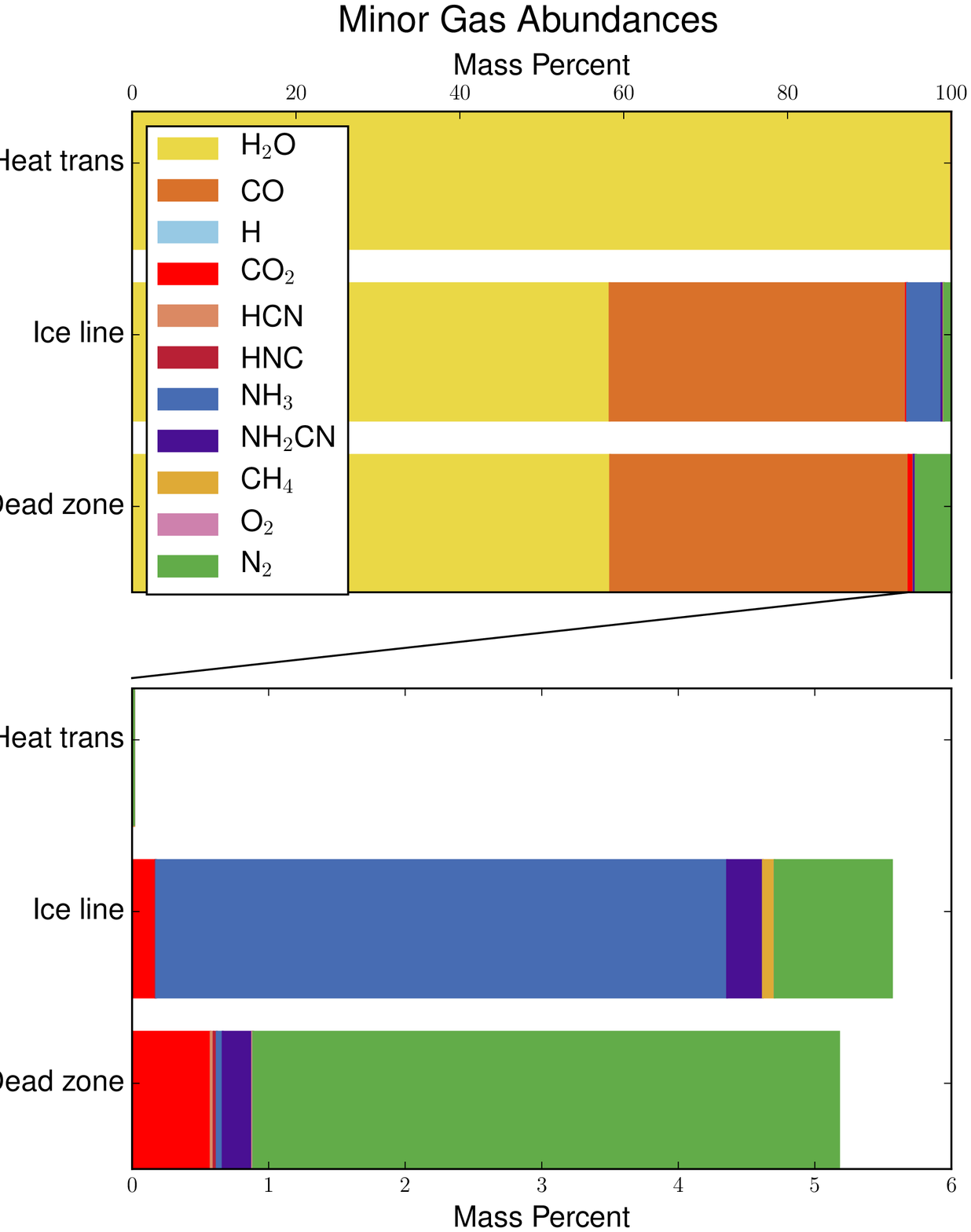}
}
\label{fig:res05}
\caption{ Resulting bulk compositions of minor gases for the low and high mass disk models. The atmospheres of each of the planets formed in these models had similar abundances (by mass) of water and CO because they accreted their atmospheres at similar positions relative to the water ice line. The nitrogen content shows some variation based on the timing of atmosphere accretion and the global temperature structure of their natal disks. }
\end{figure*}

\begin{table}
\begin{center}
\caption{Elemental ratios at $t=t_{life}$ from the low and high initial mass dust models}
\begin{tabular}{l c c c}\hline
$M_{disk,0} = 0.08$ & $C/O$ & $C/N$ \\\hline
Dead zone & $ 0.288 $ & $4.12$ \\\hline
Ice line & $ 0.288 $ &  $4.14$ \\\hline
Heat transition & $ 0.289 $ & $4.13$ \\\hline
$M_{disk,0} = 0.12$ & $C/O$ & $C/N$ \\\hline
Dead zone & $ 0.288 $ & $ 4.12 $ \\\hline
Ice line & $ 0.288 $ & $ 4.09 $ \\\hline
Heat transition & N/A & N/A \\\hline
\end{tabular}
\label{tab:res04}
\end{center}
\end{table}

In Figures \ref{fig:res05a} and \ref{fig:res05b} we show the resulting bulk chemical composition of the minor gases in each planet's atmospheres, and in Table \ref{tab:res04} we show the elemental ratios. The water and CO abundances (by mass) of each of the planets are nearly equal, attributed to the fact that each planet accreted its gas near the ice line, or at smaller radii. Within the radius of the water ice line the relative abundance of water vapour and CO gas is constant.

The abundances of the nitrogen carriers do vary between the planets, in a similar manor as in the fiducial model. As in the fiducial model, in the high mass disk model the ice line planet accretes its solid core rapidly, accreting an atmosphere before 1 Myr. Early in the disk's evolution it is warmer and the nitrogen content is generally in NH$_3$. In the lower mass disk model the initial mass accretion rate is lower and hence the gas begins cooler. Because of its initially cooler state,  nitrogen is primarily in molecular gas (N$_2$) rather than NH$_3$. This difference is then reflected in the bulk chemical composition of the water ice line planet. Additionally in the lower mass disk model the water ice line planet does not migrate very far before accreting an atmosphere, and hence it accretes its gas in a cooler region of the disk than the water ice line planets in the two other disk models.

\begin{figure*}
\centering
\includegraphics[width=\textwidth]{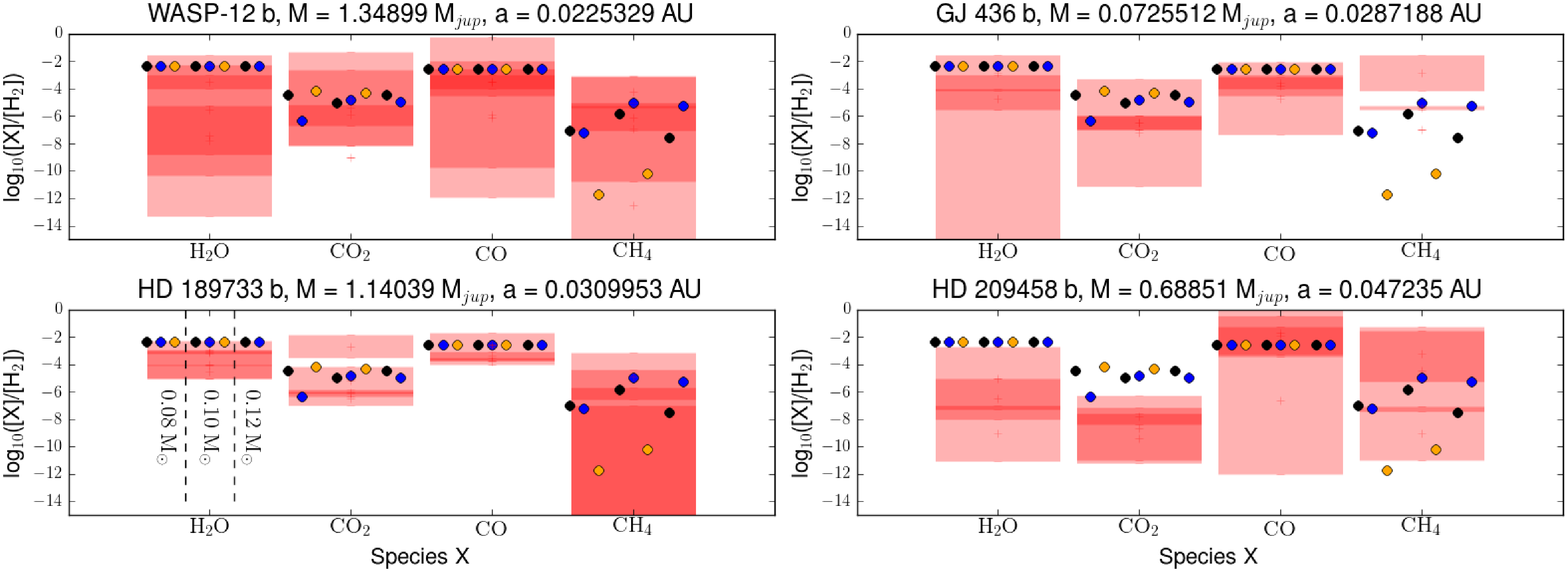}
\caption{ A comparison between the mixing ratios of molecules inferred by observations (red bars) with our theoretically derived planetary atmospheres. The three disk models with varying initial disk mass are shown, with the left three points represent the low disk mass, the middle three representing the fiducial mass disk and the right two points showing the high mass disk results. The colour of the points denote the planets formed in the dead zone (black), the water ice line (blue), and heat transition (orange) traps. The observational data comes from \citet{MadSea09,MadSea10,MigKal13,Lee12,Lin11,Line14}. }
\label{fig:res06}
\end{figure*}

In Figure \ref{fig:res06} which is one of the key results of this paper, we show the mixing ratio of H$_2$O, CO$_2$, CO, CH$_4$ for our predicted atmospheres. Additionally, we show the range of mixing ratios that are inferred by retrieval studies \citep{MadSea09,MadSea10,MigKal13,Lee12,Lin11,Line14} of atmospheric emission spectra. We see that the H$_2$O and CO content in our predicted atmospheres does not vary across our three disk models. The reason for this similarity is because each disk model is initialized by the same chemical abundance, and each planet accreted its atmosphere within the water ice line of their respective disks. These results suggest that as an observing tool, if the atmospheric mixing ratios of H$_2$O an CO are similar to the host star then this would suggest that the planet accreted its gas within the water ice line. Within the water ice line, most of the carbon and oxygen carriers are found in the gas phase \citep{Oberg11}. Since there is little to no freeze out of carbon and oxygen 
carrying 
molecules, the chemical composition of the gas is similar to the initial conditions of the disk. This result could change if the majority of carbon and oxygen come from the late accretion of planetesimals or pebbles (for eg. \cite{Mordasini16}). This late stage of accretion occurs after the evaporation of the gas disk and is not explored in this work.

Even with the same initial chemical abundances in each disk model we find small variations in the mixing ratio of CO$_2$, and large variation in the mixing ratios of CH$_4$. These variations are directly linked to the different formation histories, and are sensitive to {\it when} and {\it where} the planet accretes its gas. These variations do depend both on the initial mass of the protoplanetary disk and when the planet accreted its gas, which complicates future interpretation of observed chemical abundances in exoplanetary atmospheres.

\section{ Discussion }\label{sec:dis}

The dust component of a protoplanetary disk has important observational and theoretical effects on the structure of the disk. We have seen that the dust component of our disk models impacts the ionization structure and hence the formation history and size of planets forming in planet traps. In the context of exoplanetary atmospheres, different formation histories lead to different bulk compositions in the atmospheres of exoplanets. Understanding how differences in these bulk compositions arise could lead to a method of broadly interpreting the formation history of an exoplanet based on the observation of its atmospheric chemical content.

These interpretations are complicated by the chemical and physical evolution of the atmosphere as the planet ages, and it remains to be seen by how much these global mixing ratios are changed when the atmosphere is allowed to evolve. Of particular importance could be the interaction between the atmosphere and the core. Heavy elements in the atmosphere could be injected by out-gassing or through the erosion of the core (for ex. \cite{AliDib2017}). We leave these calculations to future work.

The fact that we form Jupiter sized planets very early ($t < 1$ Myr) in the disk lifetime has important implications on observed protoplanetary disks like HL Tau which has features that are consistent with a planet-induced gap in the dust \citep{Jin2016}. Because of its young age ($\lesssim 1$ Myr), other work has suggested that the gaps are more consistent with a change in average grain size at condensation fronts (eg. \cite{Zhang15,Okuzumi2016}). While the condensation front interpretation of disk gaps like in HL Tau may be more consistent with reality than a planet gap, we note that the concept of a Jupiter massed planet cannot, at this stage, be disregarded as it appears to be easily produced at a water ice line that is fed a substantial amount of dust by radial drift.

\subsection{ Ubiquity of H$_2$O and CO mixing ratios? }

For our three models presented here we find that each of the planets show similar ratios of H$_2$O and CO regardless of the disk model, and their C/O are similar to the ratio in the initial chemical conditions of the disk. This appears to be caused by the fact that each of the planets presented here either accreted their gas close to the location of the water ice line or at smaller radius than the location of the water ice line. The convergence of the planet traps is a feature of the particular disk models presented here, however it is not clear whether this result will stand until an exploration through parameter space can be made.

These results contrast with some observations that suggest that planets have C/O higher than their host stars \cite{Brewer16}. If we assume that the stellar photosphere represents the chemical state of the protoplanetary disk that the planets formed from, then these observations suggest that most Hot Jupiters accreted their gas beyond the water ice line. Our work suggests that the opposite is true for the case of planet formation in the presence of planet traps. Since all of our planets either accreted their gas within or very near to the water ice line, thereby accreting `pristine' gas - with the same C/O as their disk and host star. We note that a recent retrieval survey for eight hot Jupiters reports upper limits of C/O $< 0.9$ \citep{Benneke15}. These results are more consistent with ours because they imply that Hot Jupiter atmospheres tend to be oxygen rich, however their atmospheric C/O is never compared to the C/O of the host star.

A method of altering the atmosphere's C/O post-formation is through the accretion of icy planetesimals \citep{Mordasini16}. Chemically similar to comets, these planetesimals would deliver additional carbon and oxygen to the planet's atmosphere. To what extent this process would alter the observable C/O will depend on the quantity of accreted planetesimals, their chemical composition, and how deeply they travel before disintegrating \citep{Mordasini15}. Generally, oxygen is more abundant than carbon in the frozen content of comets, which would drive C/O lower than would be accreted directly from the gas disk. This would drive C/O below unity for planets which accrete gas beyond the water ice line, and below the C/O of their host star for planets which accrete their gas within the water ice line.

The tendency of planet traps to converge to the same radius in these particular runs at late times is caused by the rapid depletion of dust in the outer disk (r $>$ r$_{iceline}$). This latter feature of our model is inconsistent with sub-millimeter observations of protoplanetary disks, which have observed dust out to hundreds of AU. Such an issue as raised in CPB17 where we find that radial drift clears out a substantial amount of dust quickly in the outer disk. A potential fix for such a rapid clearing of material is dust trapping at pressure maxima \citep{Pinilla12}. Indeed in our planet formation model we assume that planet traps act as these dust traps, and we enhance the density of solids accordingly. However in our model of dust evolution these traps are not implemented. A possible outcome of these traps that more solid material is maintained in the outer disk, thereby increasing the initial formation rate of planetary cores for the planets trapped at the dead zone and heat transition.

Additionally, the location and evolution of the dead zone may have an important impact on the evolution of the dust size and density distributions. As was shown in CPB17 the dead zone suppresses fragmentation and enhanced dust settling, increasing the density of large grains and reducing the density of sub-micron grains along the midplane of the disk. In a future work we will explore a model of co-evolving gas and dust to incorporate the effects of an evolving dead zone over then entire disk lifetime.

\subsection{ Variation in CH$_4$ as a tracer of formation history? }

The carbon-carrying molecule that showed the greatest level of variation between each of our modeled planets was CH$_4$. Generally, we find that the planets which accreted their atmosphere in the colder parts of the disk showed the smallest mixing ratios for this molecule. This implies that an observation of a small CH$_4$ mixing ratio could imply that the planet formed near or outside the water ice line. We note, however that CH$_4$ tells you very little about {\it when} a planet accretes its atmosphere. As an example in the disk with initial mass of 0.08 $M_{\oplus}$ the ice line and dead zone planets accreted at very different times, but resulted with nearly identical CH$_4$ mixing ratios.

These conclusions will become complicated by the equilibrium chemistry that occurs in exoplanetary atmospheres. In equilibrium chemistry, there is a sharp transition at a temperature of $\sim 750$ K above which CH$_4$ is converted into CO (see for ex. \cite{Pignatale2011}). This transition will greatly impact our ability to observe CH$_4$ in Hot Jupiters because their equilibrium temperatures can exceed 1000 K \citep{Wakeford17}.

\ignore{
Possibly, when combined with the mixing ratio of CO$_2$, CH$_4$ could be used to trace the formation history of the planet. In the case of the 0.08 M$_\oplus$ initial mass disk, the mixing ratio of CO$_2$ for the dead zone planet is larger than the ice line planet. So we might say that given two planets, both with the same CH$_4$ mixing ratio, and one with a higher CO$_2$ mixing ratio than the other. One might say that the planet with the higher CO$_2$ mixing ratio accreted its atmosphere at a later time than the other planet, while both planets appear to have accreted its gas near the location of the water ice line. 
}

\subsection{ Nitrogen carriers as a tracer of formation history? }

With the James Webb Space Telescope (JWST) launch in 2018 we will have a new tracer of formation history, NH$_3$. The camera JWST-MIRI will study the mid-infrared which gives us the first chance to directly detect features caused by the presence of NH$_3$ in the emission spectra of atmospheres. The first possible detection of nitrogen chemistry has been recently reported by \cite{MacDonald2017}. In that work, the authors report a new retrieval technique (known as POSEIDON) which they apply to the spectra of HD 209458b. \cite{MacDonald2017} report a range of mixing ratios for NH$_3$ between 0.01 - 2.7 ppm relative to the abundance of hydrogen atoms, depending on their choice of atmospheric model. Across all of our disk models presented here, we find mixing ratios as low as $2\times 10^{-6}$ ppm, and as high as 52 ppm. These results results suggest that with current HST observations and advanced retrieval algorthims (eg. \cite{Lavie16,MacDonald2017}) we can begin to constrain the formation histories of 
observed planets with high abundances of NH$_3$.

As was pointed out in CPA16, the detection of NH$_3$ in an atmosphere might be indicative of a planet that accreted its gas early ($t < 1$ Myr) in the lifetime of the disk. A non-detection of NH$_3$ does not necessarily indicate a planet that accreted gas later, because some atmospheric evolution could have removed the molecule from the upper atmosphere. 

\section{ Conclusions }\label{sec:con}

In this work we have demonstrated that complex dust physical models can drastically change the formation history of planets forming in planet traps. With the changing formation history comes a change in {\it when} and {\it where} the planet accretes its atmosphere and hence a change in the bulk atmospheric abundances of the gas. 

Changing the surface density of dust grains throughout the disk impacts the freezing efficiency of volatiles and their formation pathways which are dependent on the presence of dust. This has lead to: \begin{itemize}
\item A retention of volatile H$_2$O and CO$_2$ when compared to the CPA16 (constant dust-to-gas ratio) chemical model at $r>r_{iceline}$
\item An enhancement of nitrogen carriers N$_2$ and NH$_3$ are found at $r<r_{iceline}$ due to formation reactions that are catalyzed by the presence of dust
\end{itemize}

The radial drift of dust results in a rapid depletion of dust in the outer parts of the disk, and a higher ionization fraction at larger radii ($r > r_{iceline}$). These higher ionization fractions change the formation history of planets forming in our model producing: \begin{itemize}
\item Hot Saturns that form at the dead zone edge
\item Earth and Super-Earth sized planets at the heat transition trap
\item Hot Jupiters and 1 AU Jupiters from the water ice line trap
\end{itemize}

We have begun to explore the available parameter space of disk initial masses in order to understand what range of molecular abundances could exist in an `early' exoplanetary atmosphere. This early study has suggested that: \begin{itemize}
\item The mixing ratio of H$_2$O and CO appear to be constant across planets that formed within the water ice line, and produce C/O that are the same as the ratio of the protoplanetary disk
\item CH$_4$ shows a large variation between different formation histories and is tied to the planet's proximity to the disk's water ice line
\item NH$_3$ could act as a measure of {\it when} the planet accretes it's gas
\begin{itemize}
\item Earlier on, the disk is warmer and NH$_3$ dominates the nitrogen carriers
\item Later the disk cools and the nitrogen is found predominately in molecular nitrogen gas
\end{itemize}
\end{itemize}

Understanding the differences in the abundance of elemental carriers is important as next generation of telescopes come online and begin to observe the emission spectra of exoplanetary atmospheres. As the library of detectable chemical species grow our models can evolve to understand where the observable diversity of chemical species arise.

\section*{ Acknowledgements }
We thank our referee, Barbara Ercolano for her insightful comments that helped to improve the paper. The work made use of the Shared Hierarchical Academic Research Computating Network (SHARCNET: www.sharcnet.ca) and Compute/Calcul Canada. A.J.C. acknowledges funding from the National Sciences and engineering Research Council (NSERC) through the Alexander Graham Bell CGS/PGS Doctoral Scholarship. R.E.P. is supported by an NSERC Discovery Grant. T.B. acknowledges support from the DFG through SPP 1833 ``Building a Habitable Earth" (KL 1469/13-1). LIC acknowledges the support of NASA through Hubble Fellowship grant HST-HF2-51356.001-A awarded by the Space Telescope Science Institute, which is operated by the Association of Universities for Research in Astronomy, Inc., for NASA, under contract NAS 5-26555. E.A.B is supported by the National Science Foundation grant AST-1514670 and AST-1344133 (INSPIRE) along with NASA XRP grant NNX16AB48G. R.E.P. also thanks the MPIA and the Institut f\"ur Theoretische 
Astrophysik (ITA) 
in the Zentrum f\"ur Astronomie Heidelberg for support during his sabbatical leave (2015/16) during the final stages of this project. A.J.C also thanks MPIA and the Institut f\"ur Theoretische 
Astrophysik (ITA) in the Zentrum f\"ur Astronomie Heidelberg for their hospitality during his 1 month stay in 2016.

\bibliography{mybib}{}
\bsp
\section*{Appendix}
\appendix
\section{ Consistent Treatment of Envelope Opacity }\label{sec:append01}

As an aside we parameterize the critical mass above which gas accretion begins, as well as gas accretion rate so that the two physical processes are assuming the same dust opacity of 3.33 cm$^2$/g in the planetary envelope. This parameterization differs from our previous work because historically the parameters are treated separately due to the fact that the opacity scaling of the critical mass and gas accretion rate change for different opacity tables. We noted in the text that the choice of parameter greatly change the formation history, and hence will have an impact on the resulting bulk chemical composition of the atmosphere.

\begin{figure}
\centering
\subfigure[Planetary formation tracks for the fiducial CPB17 model]{
	\label{fig:app01a}
	\includegraphics[width=0.5\textwidth]{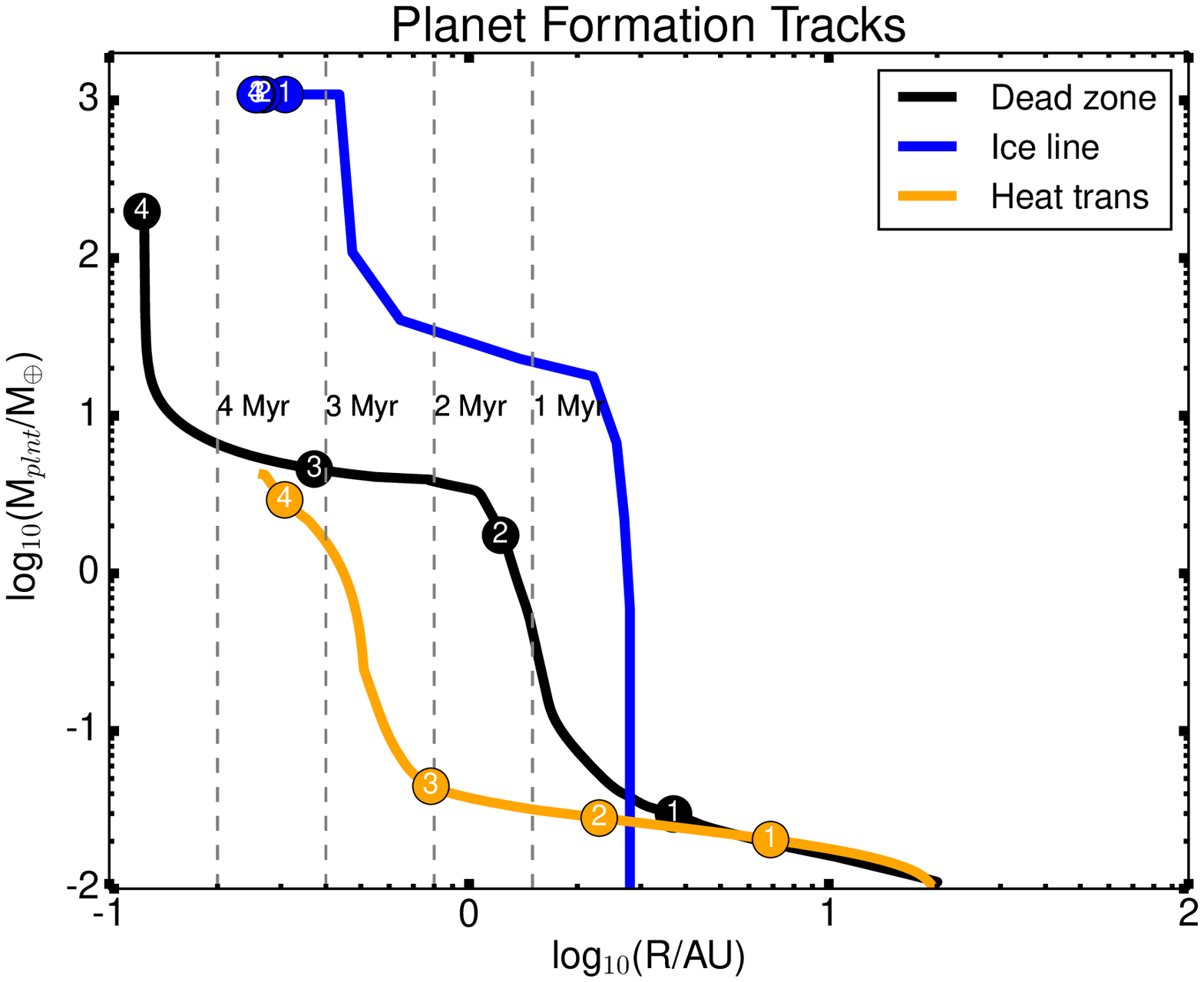}
}
\subfigure[Planetary formation tracks for CPB17 model, with $f_{crit} = 1.6$ ]{
	\label{fig:app02b}
	\includegraphics[width=0.5\textwidth]{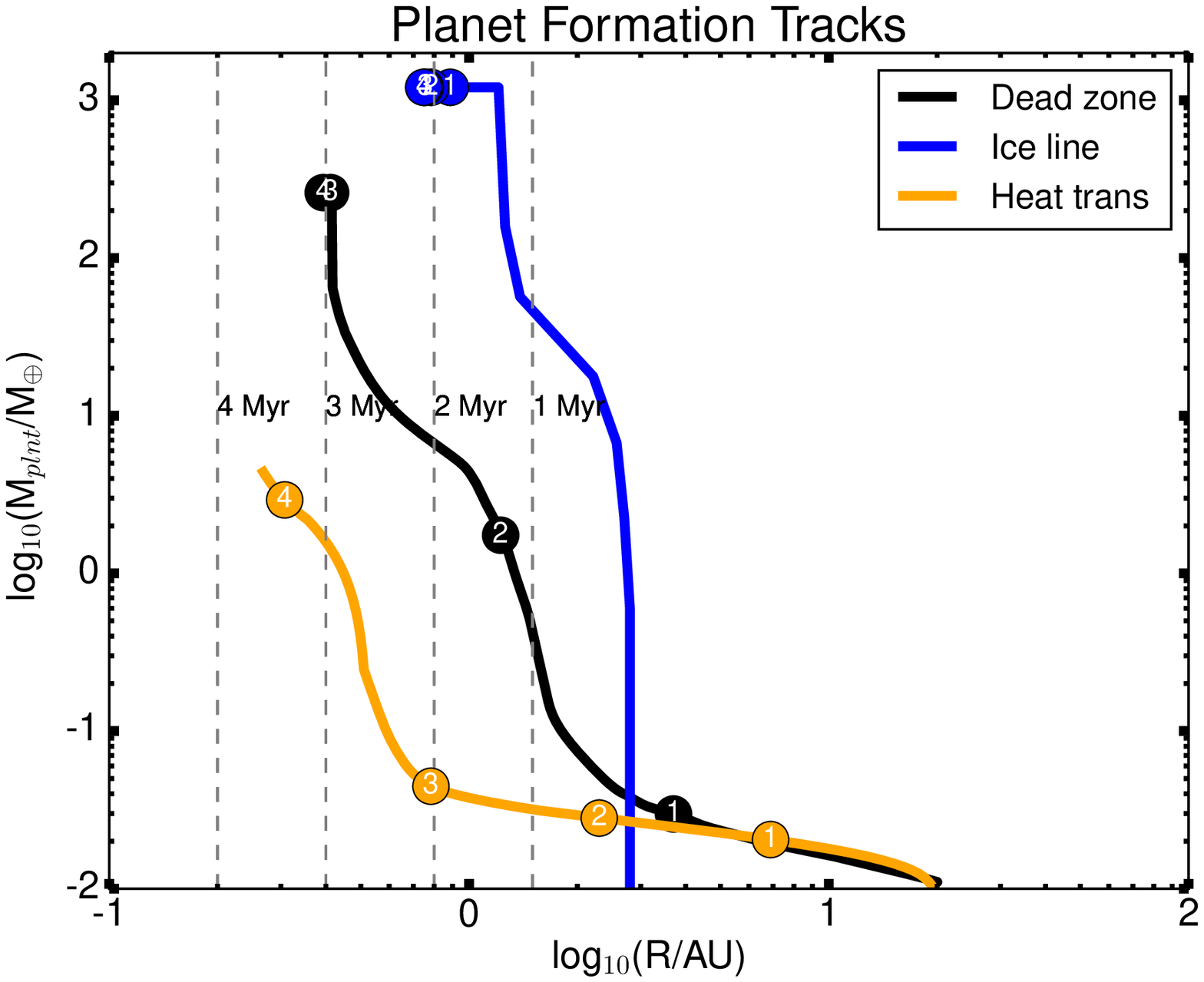}
}
\label{fig:app01}
\caption{ Comparison of the formation tracks for the fiducial CPB17 disk model with the CPB17 disk model with the planet formation parameter $f_{crit} = 1.6$. }
\end{figure}

In Figures \ref{fig:app01a} and \ref{fig:app02b} we show the formation track of the fiducial version of the CPB17 disk model and the same disk model with the parameter $f_{crit} = 1.6$. Here we see that the formation history changes drastically for the largest planets because the timing of their gas accretion changes. Importantly the core mass has to grow to a larger mass before it can begin to bring down an atmosphere. 

Because of its importance on the formation history of the planet, we will incorporate variations in the envelope opacity in our future population synthesis models.

\label{lastpage}

\end{document}